\documentclass[twocolumn]{aa}  

\usepackage{epsfig,graphicx,natbib}
\usepackage{longtable}
\usepackage{amssymb,amsmath}

\def\Tcmb{\hbox{$T_\mathrm{CMB}$}}

\def\Tex{\hbox{$T_\mathrm{ex}$}}
\def\Tkin{\hbox{$T_\mathrm{kin}$}}
\def\Ncol{\hbox{$N_\mathrm{col}$}}

\def\nH2{\hbox{$n_\mathrm{H_2}$}}
\def\kms{\hbox{km\,s$^{-1}$}}

\def\PKS1830{\hbox{PKS\,1830$-$211}}
\def\cmsq{\hbox{cm$^{-2}$}}
\def\ccm{\hbox{cm$^{-3}$}}
\def\fH2{\hbox{$f_{\rm H2}$}}
\def\Dmu{\hbox{$\Delta \mu$/$\mu$}}
\def\fc{\hbox{$f_c$}}
\def\Kmu{\hbox{$K_\mu$}}
\def\Elow{\hbox{$E_{low}$}}

\begin{document}

\title{A study of submillimeter methanol absorption toward \PKS1830:}
\subtitle{Excitation, invariance of the proton-electron mass ratio, and systematics}

\author{S.~Muller \inst{1}
\and  W.~Ubachs \inst{2}
\and  K.\,M.~Menten \inst{3}
\and  C.~Henkel \inst{3,4}
\and  N.~Kanekar \inst{5}}
\institute{Department of Space, Earth and Environment, Chalmers University of Technology, Onsala Space Observatory, SE-43992 Onsala, Sweden
\and Department of Physics and Astronomy, Vrije Universiteit, De Boelelaan 1081, 1081 HV Amsterdam, The Netherlands
\and  Max-Planck-Institut f\"ur Radioastonomie, Auf dem H\"ugel 69, D-53121 Bonn, Germany
\and  Astron. Dept., King Abdulaziz University, P.O. Box 80203, Jeddah 21589, Saudi Arabia
\and National Centre for Radio Astrophysics, Tata Institute of Fundamental Research, Pune University, Pune 411007, India}

\date{Received  / Accepted}

\titlerunning{CH$_3$OH toward \PKS1830}
\authorrunning{S. Muller et al.}

\abstract{Methanol is an important tracer to probe physical and chemical conditions in the interstellar medium of galaxies. Methanol is also the most sensitive target molecule for probing potential space-time variations of the proton-electron mass ratio, $\mu$, a dimensionless constant of nature.}
{We present an extensive study of the strongest submillimeter absorption lines of methanol (with rest frequencies between 300 and 520 GHz) in the $z=0.89$ molecular absorber toward \PKS1830, the only high-redshift object in which methanol has been detected. Our goals are to constrain the excitation of the methanol lines and to investigate the cosmological invariance of $\mu$ based on their relative kinematics.}
{We observed 14 transitions of methanol, five of the A-form and nine of the E-form, and three transitions of A-$^{13}$CH$_3$OH, with ALMA. We analyzed the line profiles with a Gaussian fitting and constructed a global line profile that is able to match all observations after allowing for variations of the source covering factor, line opacity scaling, and relative bulk velocity offsets. We explore methanol excitation by running the non local thermal equilibrium radiative transfer code RADEX on a grid of kinetic temperatures and H$_2$ volume densities.}
{Methanol absorption is detected in only one of the two lines of sight (the southwest) to \PKS1830. There, the excitation analysis points to a cool ($\sim 10-20$~K) and dense ($\sim 10^{4-5}$~\ccm) methanol gas. Under these conditions, several methanol transitions become anti-inverted, with excitation temperatures below the temperature of the cosmic microwave background. In addition, we measure an abundance ratio A/E $= 1.0 \pm 0.1$, an abundance ratio CH$_3$OH/H$_2$ $\sim 2 \times 10^{-8}$, and a $^{12}$CH$_3$OH/$^{13}$CH$_3$OH ratio $62 \pm 3$. Our analysis shows that the bulk velocities of the different transitions are primarily correlated with the observing epoch due to morphological changes in the background quasar's emission. There is a weaker correlation between bulk velocities and the lower level energies of the transitions, which could be a signature of temperature-velocity gradients in the absorbing gas. As a result, we do not find evidence for variations of $\mu$, and we estimate $\Dmu = (-1.8 \pm 1.2) \times 10^{-7}$ at 1-$\sigma$ from our multivariate linear regression.}
{We set a robust upper limit $| \Dmu | < 3.6 \times 10^{-7}$ (3$\sigma$) for the invariance of $\mu$ at a look-back time of half the present age of the Universe. Our analysis highlights that systematics need to be carefully taken into account in future radio molecular absorption studies aimed at testing \Dmu\ below the $10^{-7}$ horizon.}

\maketitle

\section{Introduction}

%\subsection{Interstellar methanol}

Methanol has a myriad of rotational transitions in the millimeter and submillimeter windows, many of which are commonly excited in the gas phase of the interstellar medium (ISM). It was first detected in space, toward the Galactic center, by \cite{bal70} using the 140-foot Green Bank telescope. The molecule is an asymmetric rotor\footnote{The molecule is almost a symmetric top, but the OH axis is slightly inclined relative to the symmetry axis of the CH$_3$ group.} and offers the possibility to determine both the kinetic temperature (\Tkin) and the volume density (\nH2) of molecular gas in the ISM; it is hence a powerful tool to probe physical conditions (e.g., \citealt{leu04}). Methanol is thought to be formed on the surface of dust grains, and then released into the ISM after thermal or nonthermal desorption (see, e.g., \citealt{vdtak00,mar05,gar07}). As a result, methanol emission is often associated with star-formation activity and shocks in the ISM. Methanol masers, in particular, have been used to classify the earliest phases in the process of massive star formation, depending on the excitation mechanisms at work (i.e., collisional excitation for class~I methanol masers versus infrared pumping for class~II masers: e.g., \citealt{bat87,men91}). Nonthermal desorption, for example icy grain mantle sputtering by cosmic ray particles, can also release methanol into the gas phase in quiescent dark clouds (e.g., \citealt{dar20}). 

Methanol has been observed in a wide range of interstellar environments in the Milky Way, with typical abundances relative to H$_2$ of $10^{-7}-10^{-6}$ in hot molecular cores (see, e.g., \citealt{men88,sch01}, for Orion-KL) and $10^{-9}-10^{-8}$ in cold dark and translucent interstellar clouds (e.g., \citealt{fri88,wal88,tur98}). On the other hand, it is not detected in the diffuse phase down to an abundance $<10^{-9}$ (\citealt{lis08}). Furthermore, methanol has been detected by its thermal and maser emission in nearby galaxies (e.g., \citealt{hen87,ell14,hum20}), and also in absorption in NGC3079 \citep{imp08} and in the $z=0.89$ absorber toward \PKS1830\ \citep{mul11}.

Methanol has a rich rotational spectrum of which the complexity reflects hindered internal rotation (torsional motion) of the OH radical group relative to the CH$_3$ frame. Various dedicated studies have been carried out to determine accurate rest frequencies (see \citealt{xu08} and references therein), and the BELGI program was developed for the analysis of the internally hindered rotation in combination with the asymmetric top structure \citep{hou94}. The many methanol lines are documented in various molecular spectroscopy databases (e.g., the Cologne Database for Molecular Spectroscopy (CDMS): \citealt{CDMS01,CDMS05,CDMS16}, and the JPL catalog: \citealt{pic98}).
 
Methanol exists in two different nuclear-spin states, depending on the net spin $I$ of the three protons of the CH$_3$ group. In A-type methanol (analogous to ortho-NH$_3$), the proton spins are parallel to each other, and $I = 3/2$. In E-type methanol (analogous to para-NH$_3$), one of the proton spins is antiparallel to the other two, and $I = 1/2$. Furthermore, E-type methanol has two degenerate forms \citep{lee73}. Consequently, the relative statistical abundance ratio of A-type to E-type is not the ratio of the spin degeneracies (i.e., 2), but unity.

%\subsection{Methanol and invariance of the fundamental constant $\mu$}

Spectroscopic observations of absorption lines toward high-redshift quasars can be used to test putative variations of the fundamental constants of physics on cosmological timescales (e.g., \citealt{sav56,bah67,uza11}). Such tests of Gyr-timescale evolution in constants like the fine structure constant $\alpha$ and the proton-electron mass ratio $\mu \equiv m_p/m_e$ are complementary to laboratory-based studies using atomic clocks (e.g., \citealt{hun14,god14,saf18}), which probe changes on timescales of years. While a variety of observational approaches have been explored using different combinations of spectral lines (e.g., \citealt{tho75,var93,dzu99,che03,kan04,fla07,jan11a}), they all consist of comparing the inferred redshift of an object as measured from different spectral lines whose frequencies have different dependencies (sensitivity coefficients) on the constant or constants of interest. For example, if the proton-electron mass ratio at the absorption redshift is different from its laboratory value today by $\Delta \mu$, then two lines with different sensitivity coefficients $K_{\mu}$ to a variation of $\mu$ would be observed to show a velocity (equivalently, redshift) offset $\Delta v$ given by
\begin{equation} \label{eq:mu}
  \Delta v/c = \Delta \Kmu \times \Dmu.
\end{equation}

The hydrogen molecule (H$_2$) is the most abundant molecular species in the Universe and its ultraviolet transitions have long been used to probe $\mu$-variation in absorbers towards high-redshift quasars \citep{tho75,var93}. H$_2$ has a strong-dipole absorption spectrum  shifted to the optical domain for high-redshift objects (e.g., \citealt{mal10,wee11}). In view of the fact that electronic transitions are probed, the sensitivity coefficients are small with typical spreads of $\Delta K_{\mu}=0.06$ \citep{var93,uba07}. The combined analysis of ten quasar systems with typically some 100 observable spectral lines, including lines pertaining to the HD isotopologue, has produced a constraint of  $|\Dmu| <5 \times 10^{-6}$ (at $3\sigma$) for redshifts in the range $z=2.0-4.2$ \citep{uba16}. However, it has been shown that systematic errors arising from the wavelength calibration of optical echelle spectrographs may limit the accuracy of such H$_2$-based measurements of $\mu$-variation (e.g., \citealt{gri10,rah13,whi15}).

Molecules that undergo quantum-tunneling motions in their electronic ground state exhibit much larger sensitivity coefficients than those of the H$_2$ electronic transitions, and therefore they offer more appealing alternatives to search for $\mu$-variation on cosmological timescales. One such molecule is ammonia, which has inversion lines with large sensitivity coefficients $\Kmu = -4.46$ \citep{fla07}. However, these lines come with the disadvantage of all having the same \Kmu\ sensitivity. Therefore, they have to be used in combination with, e.g., pure rotational lines of other species, exhibiting $\Kmu = -1$, to constrain \Dmu. This approach has been used to compare NH$_3$ inversion lines to HCN and HCO$^+$ lines (at $z_{\rm abs}=0.68$ toward the quasar B\,0218+357; \citealt{mur08}), to HC$_3$N lines (at $z_{\rm abs}=0.89$, toward the quasar \PKS1830; \citealt{hen09}), and to CS and H$_2$CO lines (again at $z_{\rm abs}=0.68$ toward B\,0218+357; \citealt{kan11}). From these studies, the best current ($3\sigma$) upper limits are  $| \Dmu | < 3.6 \times 10^{-7}$ from $z=0.68$ to today, and $| \Dmu | < 1 \times 10^{-6}$ from $z=0.89$ to today, the latter corresponding to a look-back time of about half the present age of the Universe. However, comparing the redshifts of lines from different molecular species introduces the potential problem of chemical (and kinematic) segregation within the molecular cloud, resulting in possible systematic errors in these analyses.

The above systematic uncertainties may be removed by using spectral lines of a single molecular species that {\em (i)}~ have transitions with different sensitivity coefficients $K_{\mu}$; {\em (ii) \textit{\emph{have}}}  $\Kmu$ larger than for H$_2$ and possibly as high as for NH$_3$, or even higher; {\em (iii)} are relatively abundant in the ISM; and {\em (iv)} are easy to observe with a sufficient signal-to-noise ratio (S/N), even in cosmological sources. At present, two species, hydroxyl (OH) and methanol (CH$_3$OH), satisfy all the above criteria; however, the OH $\Lambda$-doubled line frequencies depend on both $\alpha$ and $\mu$, and hence cannot be used to probe changes in $\mu$ alone \citep{che03}. In the case of methanol, whose line frequencies depend on $\mu$ alone, the extreme $\mu$-sensitivity is connected to the quantum tunneling of the OH-group through the threefold internal barrier spanned by the CH$_3$ group. The spread in sensitivity coefficients, \Kmu, calculated by \cite{jan11a,jan11b} and \cite{lev11}, is up to several orders of magnitude larger than for H$_2$, and in some cases, larger than the $\Delta \Kmu$ between inversion lines of NH$_3$ and other rotational lines. These features make methanol the most sensitive probe of $\mu$-variation among known interstellar molecules.

%\subsection{Methanol and \PKS1830}

While methanol is abundant in the ISM and has already been the target of studies probing spatial variations in $\mu$ in the Milky Way \citep{lev11,ell11,dap17}, the molecular absorber at $z=0.89$ toward \PKS1830\ is currently the only object at cosmological distances in which methanol has been detected \citep{mul11}. The methanol lines in \PKS1830\ have hence been used by a number of later works to probe cosmological evolution in $\mu$ \citep{ell12,bag13a,bag13b,kan15}. Indeed, this absorber has yielded the most stringent constraint on changes in any constant on Gyr timescales, $| \Dmu | < 6 \times 10^{-7}$ (at $3\sigma$ significance; \citealt{kan15}), over $\sim 7.5$~Gyr.

PKS\,1830$-$211, at a redshift $z=2.5$ \citep{lid99}, is a blazar that is strongly lensed by an intervening galaxy at $z=0.89$ \citep{wik96}, which appears as a weak, nearly face-on spiral in a 814\,\r{A} image made with the Hubble Space Telescope \citep{win02}. At millimeter wavelengths, the blazar emission is dominated by two bright and compact images \citep{fry97,mul20b} separated by $\sim 1\arcsec$, hereafter the northeast and southwest images. At centimeter wavelengths, the two images are embedded in a nearly complete Einstein ring \citep{sub90,jau91,nai93}. The Einstein ring and compact images have different spectral indices, such that \PKS1830's morphology gradually changes with frequency. The blazar is active throughout the electromagnetic spectrum, for example at millimeter and submillimeter wavelengths (e.g., \citealt{mar13,mar20,mar19}), in the radio-centimeter domain (e.g., \citealt{jin03,all17}), and in gamma rays (e.g., \citealt{abd15}).

The molecular gas in the $z=0.89$ lensing galaxy causes a remarkable absorption spectrum, with more than 60 species detected to date in the line of sight to the southwest image of \PKS1830\ (e.g., \citealt{mul11,mul14a,ter20}), including methanol \citep{mul11,bag13a}. The continuum emission associated with the southwest image has a size of only a fraction of a milli-arcsecond at millimeter wavelengths \citep{gui99}, corresponding to a pencil-beam illumination of the order of $\lesssim 1$~parsec in the plane of the $z=0.89$ absorber. This, along with the intrinsic activity of the blazar, the geometry of the system (with the blazar's jet pointed almost exactly in our direction), and the magnification due to gravitational lensing, leads to very special  conditions that are conducive to temporal variations of the molecular absorption spectra in the millimeter domain \citep{mul08,sch15}. This effect prevents the robust kinematical comparison of spectra obtained at widely different epochs and hampers studies to constrain $\mu$-variations \citep{bag13b}. We note, however, that the spectra of H\,I and OH at radio-centimeter wavelengths have been remarkably stable for more than 20 years \citep{che99,all17,com21}, perhaps due to the facts that the continuum illumination encompasses a much larger region here than at shorter wavelengths (due to the Einstein ring) and that the H\,I/OH gas exhibits a less clumpy distribution than observed for more complex molecules.

While temporal variations of the absorption profiles have been identified as the major systematics for constraining $\mu$-invariance with methanol lines toward \PKS1830, there is also evidence for inhomogeneities in the absorbing gas, with temperature and density gradients \citep{sat13,bag13b,sch15,mul20a}, possibly affecting the comparison of lines with different excitations. Therefore, such systematics also need to be addressed by new studies of $\mu$ variability. Attempts to spatially resolve the methanol absorption have been undertaken with long-baseline interferometry \citep{mar17}, but with little success so far, due to severe limitations of sensitivity and dynamic range.

This aims of this work is a comprehensive study of the strongest submillimeter lines of methanol toward \PKS1830. The paper is divided into the following sections: observations are described in Sect.\,\ref{sec:obs}; overall results are given in Sect.\,\ref{sec:results}; we analyze the line profile in Sect.\,\ref{sec:GlobalProfile}; we constrain the excitation of the methanol gas in Sect.\,\ref{sec:excitation}; and we investigate the invariance of $\mu$ in Sect.\,\ref{sec:mu}, with a particular attention given to systematics. A summary is given in Sect.\,\ref{sec:summary}.

\section{Observations} \label{sec:obs}

\begin{table*}[h]
\caption{Summary of the observations.}
\label{tab:obsdata}
\begin{center} \begin{tabular}{ccccccccccccc}
\hline \hline
Tuning & Freq. $^{(a)}$ & Date &  N$_{\rm ant}$  & B$_{\rm min}$ $^{(c)}$ & B$_{\rm max}$ $^{(c)}$ & PWV $^{(d)}$ & t$_{\rm on}$ $^{(e)}$ & $\delta v$ $^{(f)}$ & $\sigma$ $^{(g)}$ & Beam & \\
       & (GHz) &      &  $^{(b)}$      & (m)         & (km)         & (mm) & (min)      & (\kms)     & ($10^{-3}$)     & (mas) & \\ 
\hline
B4a & 154.9 & 2019 Jul 10 & 45 & 138 & 13.9 & 2.1 &  7:36 & 1.0 & 1.4 & 50 &  \\ % X7325 MJD=674.062 &
B6  & 230.7 & 2019 Jul 11 & 44 & 111 & 12.6 & 1.2 & 24:52 & 1.2 & 1.2 & 30 &  \\ % X14bc MJD=675.243 &
B5  & 181.6 & 2019 Jul 28 & 43 &  92 &  8.5 & 0.6 & 17:14 & 1.6 & 1.7 & 60 &  \\ % X1aa8 MJD=692.045 &
B7  & 283.4 & 2019 Jul 28 & 42 &  92 &  8.5 & 0.6 & 18:49 & 1.1 & 1.1 & 30 &   \\ % X1b99 MJD=692.079 &
B4b & 154.9 & 2019 Aug 17 & 47 &  41 &  3.6 & 1.0 &  7:37 & 1.0 & 1.0 & 180 & \\ % X21df MJD=712.126 &
\hline
\end{tabular} \end{center}
\tablefoot{$(a)$ Frequency of the local oscillator between the lower and upper sidebands;
  $(b)$ number of antennas in the array;
  $(c)$ B$_{\rm min}$, B$_{\rm max}$: minimum and maximum projected baselines, respectively;
  $(d)$ amount of precipitable water vapor in the atmosphere;
  $(e)$ total on-source integration time;
  $(f)$ velocity resolution, after Hanning smoothing;
  $(g)$ rms noise normalized to the continuum level.
}
\end{table*}

The observations were performed with the Atacama Large Millimeter/submillimeter Array (ALMA) between 2019 July and August (ALMA project code 2018.1.00051.S.) and are summarized in Table~\ref{tab:obsdata}.

In 2019 July, a total of four different tunings were observed, in bands~4 (B4, $\sim 150$~GHz), 5 (B5, $\sim 180$~GHz), 6 (B6, $\sim 230$~GHz), and 7 (B7, $\sim 280$~GHz), to cover all the targeted redshifted methanol lines. During this time, the array was in extended configurations, with the longest baselines up to 8--14~km providing synthesized beams between 30 and 60~mas. The B4 and B6 tunings were observed close together in time, on July 10 and 11. The B5 and B7 tunings were observed consecutively on July 28. The B4 tuning was reobserved with exactly the same setup on 2019 August 17 to allow us to test potential temporal variations of the absorption profile over the duration of our observational campaign. At this time, the array was in a more compact configuration, with a maximum baseline $<4$~km, yielding a synthesized beam of $0.18\arcsec$.

For the B5, B6, and B7 tunings, the correlator was configured to cover four spectral windows, each of 1.875~GHz in width and with a spectral resolution of $\sim 1$~MHz (after Hanning smoothing). For the B4 tuning, we configured three spectral windows of 0.9375~GHz with a narrower resolution of 0.56~MHz, while the remaining fourth spectral window, not covering any methanol lines, was set with a width of 1.875~GHz and a resolution of $\sim 1$~MHz. The resulting velocity resolutions are in the range 1.0--1.6~\kms\ after Hanning smoothing, for all methanol lines. The observational details are summarized in Table~\ref{tab:obsdata}.

The data calibration was done within the CASA\footnote{http://casa.nrao.edu/} package, following standard procedures. The bandpass responses of the antennas were determined from observations of the bright quasar J\,1924$-$2914, which was also used to calibrate the flux density scale. According to ALMA observatory flux monitoring data, we expect a flux scale accuracy better than 5\% at bands 4, 5, and 6, and better than 10\% at band~7.

After the standard gain calibration using the quasar J\,1832$-$2039, the phase solutions were further self-calibrated on the continuum of \PKS1830, applying phase-only gain corrections to every 6~s integration. Spectra were converted from the topocentric to LSRK\footnote{Local Standard of Rest with the kinematic (radio) definition (J2000) based on an average velocity of stars in the solar neighborhood.} reference frame with the CASA task {\texttt mstransform}. The conversion between LSRK and the heliocentric frame, which we chose as the velocity reference for easier comparison with previous work, is $v_{\rm Helio} = v_{\rm LSRK} - 12.4$~\kms. The final spectra were extracted using the CASA-Python task UVMultiFit \citep{mar14}, by fitting a model of two point sources (corresponding to the northeast and southwest core images of \PKS1830) to the interferometric visibilities.

We note that the third lensed image of \PKS1830, located about half-way between the northeast and southwest images with a flux density roughly 140 times weaker than that of the northeast source, as well as the weak extensions from the northeast and southwest images, are all detected from this same ALMA dataset \citep{mul20b}. However, these have a negligible effect on the absorption spectra extracted with a two point-source model and have hence been ignored here.

\section{Results} \label{sec:results}

\begin{figure}[h!] \begin{center}
\includegraphics[width=8.8cm]{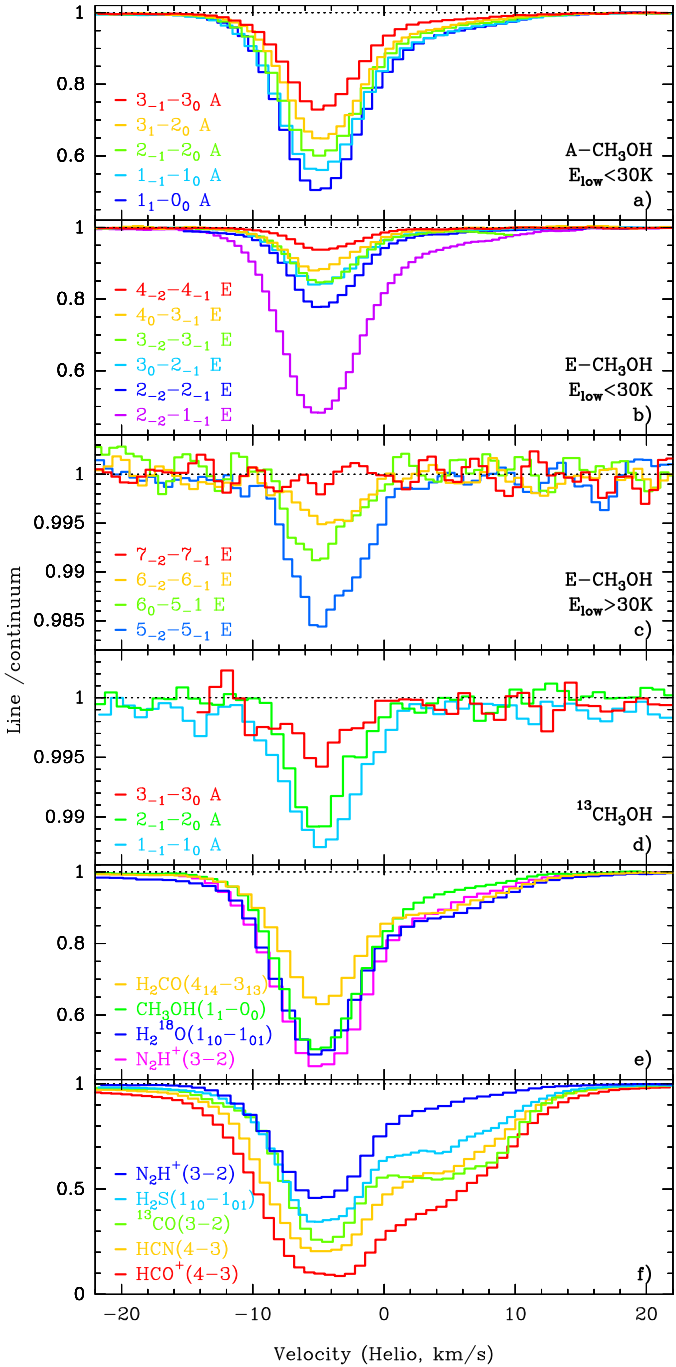}
\caption{ALMA spectra of methanol and several other species observed toward the southwest image of \PKS1830\ during our campaign (July-August 2019). For lines in B4, the spectra were averaged over both B4a and B4b observations (see Fig.\,\ref{fig:DiffB4ab} for time variability). To avoid crowded plots and large spreads in absorption depths, the lines from A-CH$_3$OH (a), E-CH$_3$OH (b,c), and $^{13}$CH$_3$OH (d) are separated, and they are also split according to low-level energy $\Elow < 30$~K (b) and $\Elow > 30$~K (c) for E-CH$_3$OH. For comparison, some other species are shown in (e), for intermediate opacity lines, and (f), for lines with the highest opacity.}
\label{fig:spec}
\end{center} \end{figure}

\subsection{Description of the spectra}

The absorption profiles of the methanol lines that we detected toward the southwest image of \PKS1830 are presented in Fig.\,\ref{fig:spec}, together with those of some other species. We detected five transitions from A-CH$_3$OH and nine from the E-form. These transitions are indicated on the methanol A- and E-state energy diagrams in Fig.\,\ref{fig:Ediag} and their spectroscopic parameters are given in Table~\ref{tab:line-methanol}. None of the methanol lines were detected in emission, implying an absence of inversion and maser effects in all measured transitions. The penultimate column of Table~\ref{tab:obsdata} lists the rms noise (normalized to the continuum level) for each band, at velocity resolutions of $\sim 1-1.6$~km/s.

At first glance, all methanol lines seem to have similar profiles, scaled by the respective line absorption strengths. The deepest absorption comes from the ground state transitions $1_1-0_0$ of A-CH$_3$OH and $2_{-2}-1_{-1}$ of E-CH$_3$OH, reaching an absorption depth of nearly 50\% of the continuum level. The weakest detected transition, $6_{-2}-6_{-1}$~E, has a lower-level energy $\Elow = 54$~K with respect to the ground state of A-CH$_3$OH, which is 7.90~K lower than that of E-CH$_3$OH. The next E-CH$_3$OH transition, $7_{-2}-7_{-1}$~E with $\Elow = 70$~K, is not detected, with an rms noise level of $1.1 \times 10^{-3}$ of the continuum level. Among the observed A-CH$_3$OH transitions, none has \Elow\ higher than 30~K.

Except for the $2_1-1_0$~A transition (rest frequency of 398.446920~GHz, redshifted to $\sim 211$~GHz, which we skipped to limit the number of tunings and observing time, and because it has properties redundant with those of other targeted lines in terms of \Elow\ and \Kmu), we expect our survey of the methanol lines in the $z=0.89$ absorber to be complete down to a peak opacity $\tau =0.1$ across all current ALMA bands (bands 3 to 10).

We also detect, for the first time toward \PKS1830, three transitions of the $^{13}$CH$_3$OH isotopologue (Fig.\,\ref{fig:spec}d). These transitions correspond to the same series $J_{-1}-J_0$ with $J= 1,2,3$ as for the main $^{12}$CH$_3$OH isotopologue in B4. We determine the $^{12}$CH$_3$OH/$^{13}$CH$_3$OH ratio in Sect.\,\ref{sec:C1213}.

As observed before (e.g., \citealt{mul11,ell12,bag13b,kan15,mar17,mul20a}), the methanol absorption has a main component peaking at a velocity\footnote{All velocities in this work refer to the heliocentric frame, using a redshift $z=0.88582$.} $v \sim -5$~\kms. The ALMA spectra clearly reveal an additional shallow absorption feature between $v = +4$~\kms\ and $+10$~\kms, which reveals itself in the new data either because of the significantly higher S/N or temporal variations in the line profile. For the methanol lines, the absorption depth of this side component is not as deep, relative to the main component,  as for transitions of H$_2^{18}$O, H$_2$CO, and N$_2$H$^+$, observed simultaneously, although these species have similar or even lower depths near $v=-5$~\kms\ (Fig.\,\ref{fig:spec}e). This suggests a relative enhancement of methanol absorption in the $v=-5$~\kms\ component.

Among all species observed in this campaign, the HCO$^+$~($4-3$) line shows the deepest absorption (Fig.\,\ref{fig:spec}f), and we infer that its continuum-covering factor is larger than 90\% for the $v \sim -5$~\kms\ component.
We discuss the methanol spectra further in Sect.\,\ref{sec:GlobalProfile}, where we construct a global absorption profile for all methanol lines.

\subsection{Time variability of the quasar and of the absorption lines}

Our observing campaign was designed to allow us to address the time variability of \PKS1830\ as a systematic. For this reason, we requested all observations be performed within the shortest possible time interval, depending on the dynamical scheduling constraints of ALMA. To assess potential temporal variations, we requested a second B4 tuning as the final observation of our campaign. In this section, we investigate the intrinsic variability of the quasar and the evolution of the line profile within the time span of our observations.

We note that \PKS1830\ endured a period of strong radio and gamma flaring activity about three months before our observations but had returned to an apparent quiescent stage by the time of our campaign (see, e.g., \citealt{mar20,abh21}).

\subsubsection{Quasar activity}

\begin{figure*}[h] \begin{center}
\includegraphics[width=\textwidth]{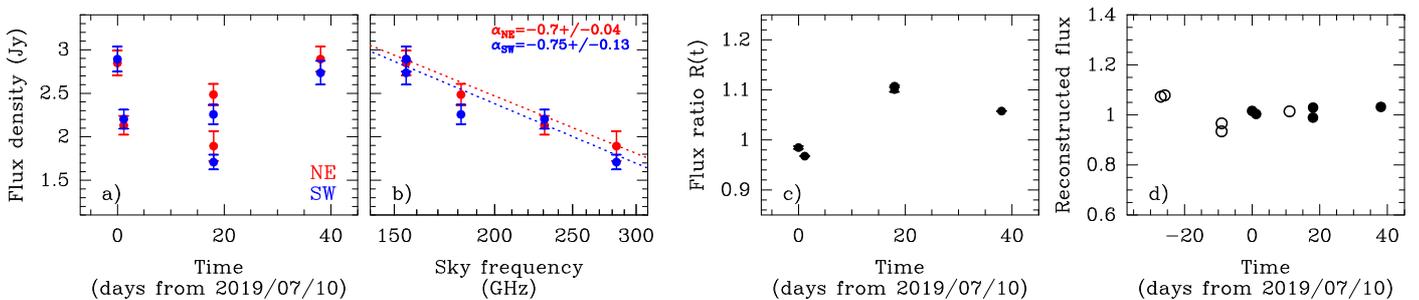}
\caption{Temporal variations of the continuum emission of \PKS1830.
  (a) Evolution of the flux density of the northeast (red) and southwest (blue) lensed images during our ALMA campaign.
  (b) Spectral energy distribution (the frequency axis is on a logarithmic scale) of the two images (i.e., same data points as for (a) but now displayed as function of frequency), with the best fit of a $F = F_0(\nu/\nu_0)^{\alpha}$ relationship indicated in dotted lines for each image.
  (c) Evolution of the instantaneous flux ratio $R(t)=F_{NE}/F_{SW}$.
  (d) Reconstructed flux density history (normalized by a flux $F=2.87$~Jy at a frequency of 150~GHz and adopting a spectral index $\alpha=-0.7$). The filled black dots correspond to datapoints from the northeast image, while the empty circles correspond to that from the southwest image, delayed in time by 27~days and rescaled adopting a relative magnification factor of 1.04 between the two images.}
\label{fig:contvar}
\end{center} \end{figure*}

We used the continuum data to infer the quasar activity during our ALMA campaign. The evolution of the flux densities and flux ratios of the two northeast and southwest lensed images is shown in Fig.\,\ref{fig:contvar}. The flux variations of the two images are delayed in time by $\sim 20-30$~days with northeast leading (see, e.g., \citealt{mul20b} and references therein), but our sparse data cannot place strong constraints on the time delay. Nevertheless, this is not critical for our study, and we used a time delay of $27$~days to reconstruct the intrinsic light curve of the quasar. The spectral energy distribution in Fig.\,\ref{fig:contvar}b indicates that the spectral index $\alpha$ (defined by the relationship $F=F_0 \times (\nu/\nu_0)^{\alpha}$) is close to $-0.7$. Then, we reconstructed the light curve of the quasar's northeast image (Fig.\,\ref{fig:contvar}d) by taking the flux density points of the northeast image corrected for the spectral index, and those of the southwest image shifted in time and corrected for spectral index and the relative magnification factor. With those additional points, we see that it is difficult to imagine large flux variations over short periods of less than $\sim 10$~days, and we thus assume that the light curve is smooth after the start of our observations. A relative magnification factor of 1.04 would minimize the flux dispersion of points between July 10 and 28, leaving only minor flux density variations during our ALMA campaign, of at most$  5$\%.
The flux ratios are within the 1--2 range of past measurements (e.g., \citealt{vanomm95,mul08,mar13,mar19,mar20}).

\subsubsection{Variability of the absorption profile} \label{sec:linevar}

\begin{figure}[h] \begin{center}
\includegraphics[width=8.8cm]{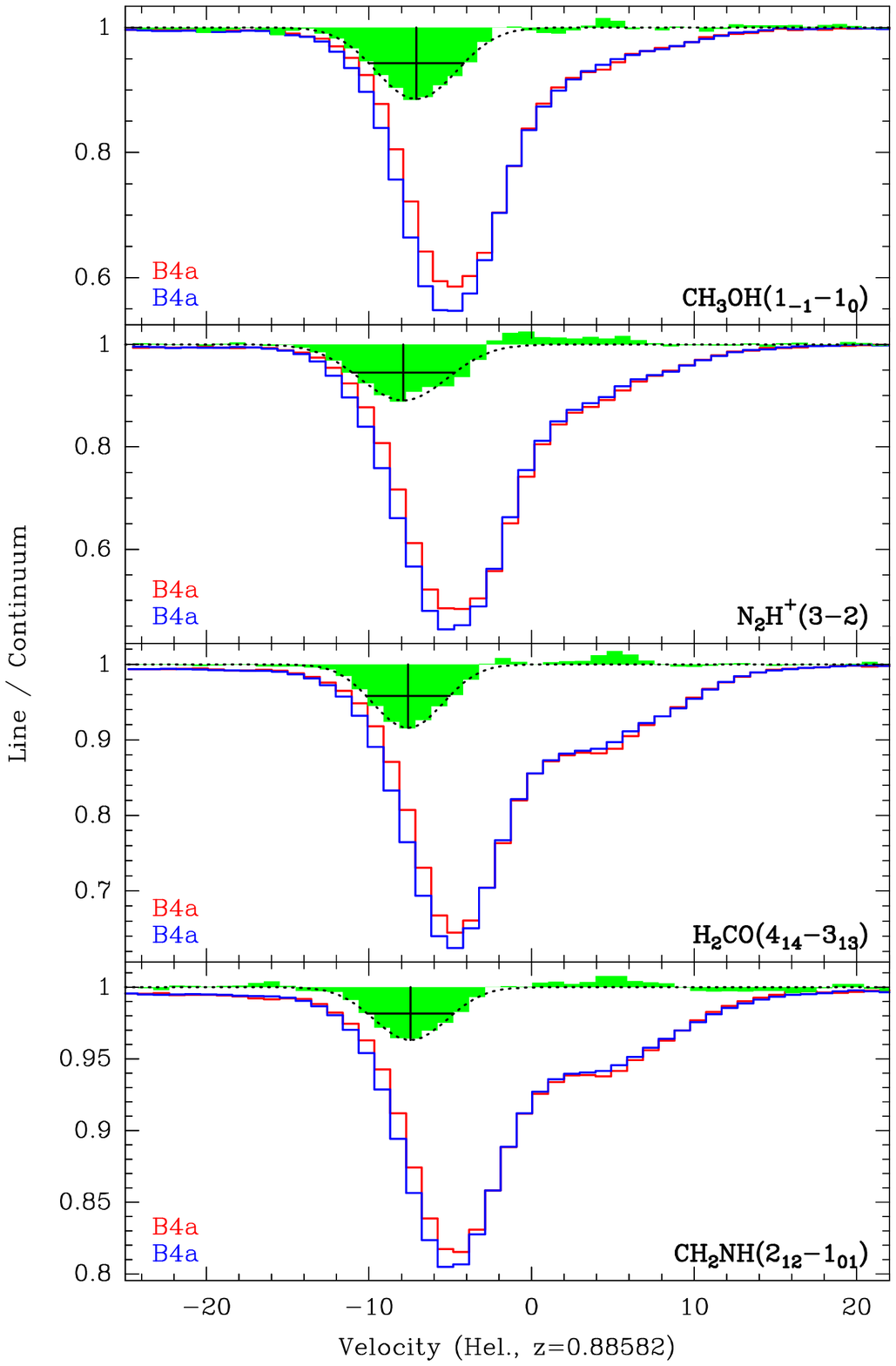}
\caption{Spectra of lines from CH$_3$OH, N$_2$H$^+$, H$_2$CO, and CH$_2$NH, observed simultaneously on 2019 July 10 (B4a, in red) and August 17 (B4b, in blue). The difference spectra (B4b--B4a) are shown in green, multiplied by a factor two and shifted to unity. The best Gaussian fit to each of the difference spectra is overlaid as a dotted line, the cross indicating the velocity centroid, amplitude, and full-width at half maximum of the Gaussian.}
\label{fig:DiffB4ab}
\end{center} \end{figure}

Since we reobserved the same tuning (B4) about one month apart, at the beginning and end of our methanol campaign, we were able to investigate the variability of the absorption profile of the different observed species. The difference spectra are shown in Fig.\,\ref{fig:DiffB4ab} for lines of CH$_3$OH, N$_2$H$^+$, H$_2$CO, and CH$_2$NH. The profiles are in excellent agreement, except for a Gaussian-like feature in the difference spectra, at a velocity of $-7.5$~\kms\ and with a full width at half maximum (FWHM) of $\sim 6$~\kms\ (i.e., in the blue wing of the main $v=-5$~\kms\ velocity component), which is common to all species. This feature corresponds to a change in integrated opacity within the velocity range $-13$~\kms\ to $-2$~\kms\ of less than 10\%.

Such a variation, limited to a short range of velocities, shows evidence of small-scale and short-timescale structural changes in the morphology of the quasar\footnote{Such changes have actually been observed with Very Long Baseline Interferometry (see e.g., \cite{gar97}).}, and suggests as well that the different velocity components are likely to have different covering factors.

The peculiar variability of the main $-5$~\kms\ velocity component of the methanol absorption profile was already noticed by \cite{mul20a}, with variations of a factor of two within a few months (between 2019 April and July). This, plus its chemical properties, led \cite{mul20a} to conclude that this velocity component may be associated with a compact dark cloud similar to those in the Milky Way.

\subsection{Non-detection of methanol on the northeast line of sight}

For completeness, we present the stacked spectrum corresponding to the weighted average of all A- and E-CH$_3$OH transitions with $\Elow < 30$~K toward the northeast image in Fig.\,\ref{fig:specNE}. Each observation was resampled to the same velocity grid with a velocity resolution $\delta v = 2$~\kms\ and we used the squared inverse of the rms noise levels in Tab.\,\ref{tab:obsdata} as relative weights. The rms noise of the stacked spectrum is $\sigma_{\tau} = 2.9 \times 10^{-4}$ of the normalized continuum level with $\delta v = 2$~\kms. Methanol remains undetected along this line of sight, which is known to contain more diffuse interstellar material than the southwest absorbing column (e.g., \citealt{mul14b,mul17}).

We calculated an upper limit of the methanol northeast column density from this non-detection. First, we determined the upper limit on the integrated opacity as $3\sigma_{\tau}\sqrt{\delta v \Delta v}$, where we assumed a line width $\Delta v = 20$~\kms\ for the absorption near $v=-150$~\kms\ (e.g., \citealt{mul14a}). For the ground-state transition of A-CH$_3$OH (350.905~GHz rest frequency and one of the strongest absorption lines in our campaign), we calculated a conversion factor $\alpha = 2.2 \times 10^{13}$~\cmsq\,km$^{-1}$\,s between integrated opacity and column density ($N_{col} = \alpha \int \tau dv$) under conditions of local thermodynamic equilibrium (LTE) with $\Tex=\Tcmb=5.14$~K. With a ratio A/E~$=1$, we then estimated a total methanol column density $< 2.4 \times 10^{11}$~\cmsq\ ($3\sigma$ upper limit). With an H$_2$ column density of $1 \times 10^{21}$~\cmsq\ \citep{mul14a}, this upper limit corresponds to an abundance ratio CH$_3$OH/H$_2$~$< 2.4 \times 10^{-10}$, which is well below that in the southwestern line of sight (see Sect.\,\ref{sec:abundance}) and consistent with the upper limit on methanol relative abundance in Galactic diffuse gas ($<10^{-9}$, \citealt{lis08}).

\section{Construction of a global line opacity profile for CH$_3$OH} \label{sec:GlobalProfile}

One of our main goals was to seek for potential velocity offsets between the different methanol lines in order to constrain the invariance of $\mu$. We therefore need to adopt a model to compare the different transitions. A velocity centroid analysis would be a first order option, but one that is biased by the combined effects of the asymmetric absorption profile and opacity distortion of the spectra. Similarly, a cross-correlation analysis (e.g., similar to the approach followed by \citealt{kan15}) would not be suitable for the ALMA spectra since a number of the CH$_3$OH lines of Fig.\,\ref{fig:spec} are not optically thin. This is why we turned to a fit of a single common absorption profile, in terms of optical depths, to all methanol transitions toward the southwest image of \PKS1830.

We started our fitting exercise by forcing the background source covering factor (see Appendix\,\ref{app:fc}) $\fc = 1$ for all spectra, but quickly realized that there were some tensions in the fit, as the deepest absorption could not be reproduced correctly and was systematically underestimated. Thus, we let \fc\ be a free parameter. Although we see evidence for a varying \fc\ across the absorption velocity interval, we adopted a constant \fc\ as  function of velocity because the temporal variations of the absorption profile have a small amplitude over the $\sim 1$-month time span of our observations, and they are limited to a small range of velocities (Sect.\,\ref{sec:linevar}). In addition, we wanted to minimize the complexity of the fit and number of fit parameters. However, we allowed \fc\ to vary between the different tunings to account for potential variations with the observing epoch and frequency.

We chose to model the common line profile as a sum of Gaussian velocity components $G_{ki}$ with velocity centroids $v_k$, FWHM $\Delta v_k$, and integrated opacities $\int \tau dv_k$, scaled by an opacity factor $\gamma_i$ and with a free bulk velocity $v_i$ for each transition $i$ in a joint fit. Therefore, our model of the absorption profile $S_j$ for all transitions $i$ in tuning $B_j$, is described as follows:
\begin{equation} \label{eq:commonprofile}
S_j = 1 - f_{c}(B_j) \times \left [ 1- \exp{ \left ( - \sum_i \gamma_i \sum_{k} G_{ki} \right )} \right ],
\end{equation}
\noindent where, in practice, we fixed the first Gaussian as $G_{1i}= G_1(v_i, \Delta v_1, (\int \tau dv)_1)$, to which we tied the next one as $G_{2i}= G_2(v_i+\delta v_{12}, \Delta v_2, (\int \tau dv)_2)$, with $\delta v_{12}$ the velocity separation between the first and second Gaussian. Hence, this velocity separation and the width and integrated area of each Gaussian component are forced to be constants for all lines through the fit. The $v_i$ and $\gamma_i$ are free parameters for each line. We chose the ground-state transition of A-CH$_3$OH (rest frequency 350.905~GHz, redshifted into band~5) as a reference for the normalization for the line profile ($\gamma=1$), given that, as one of the brightest lines, it has a high S/N and it was observed roughly in the middle of our campaign. We checked that changing the reference line to another with comparable S/N did not change significantly the resulting covering factors and characteristics of the line profile.

For the fit, we used the Python scipy.optimize least-squares routine based on the Levenberg-Marquardt algorithm. The uncertainties are derived as the square root of the diagonal elements of the covariance matrix multiplied by the reduced $\chi^2$. Hence, we assume that errors follow a Gaussian distribution, that the parameters are not correlated, and that the model is a good representation of the data.

\begin{figure*}[ht!] \begin{center}
\includegraphics[width=\textwidth]{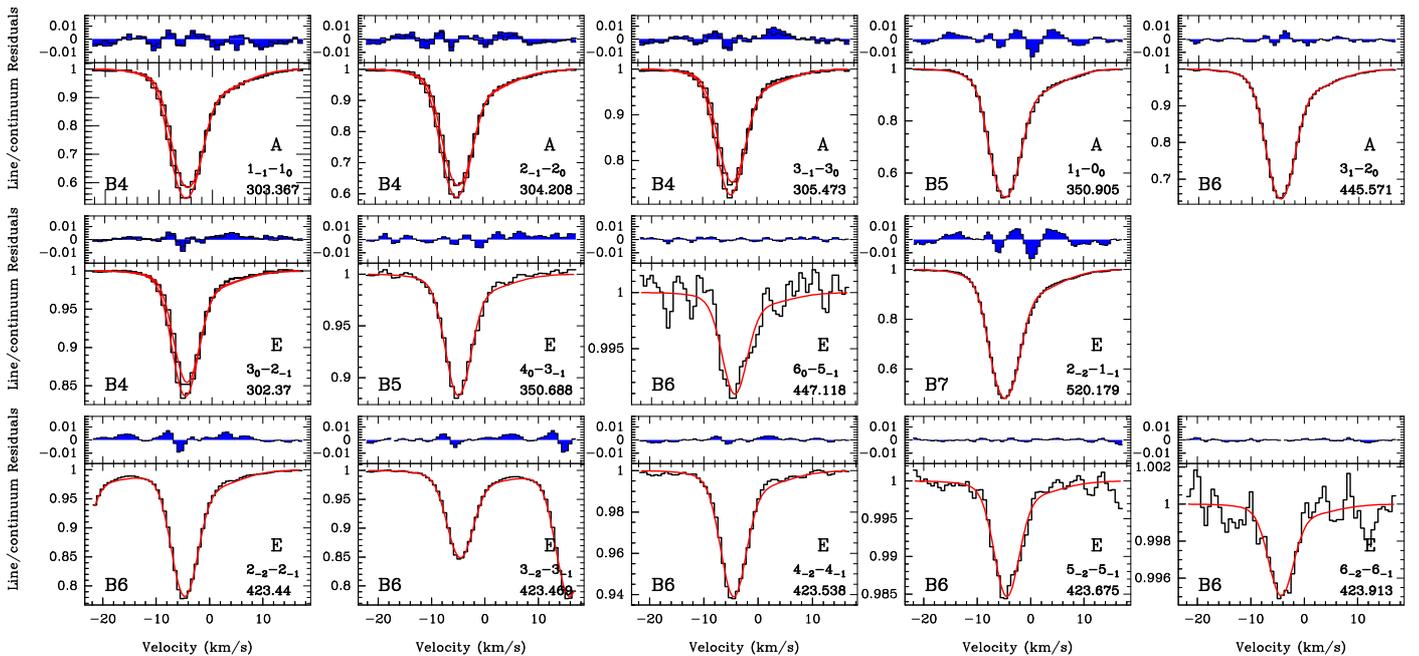}
\caption{Spectra of methanol transitions observed with ALMA toward the southwest image of \PKS1830, with best fits of the common two-Gaussian velocity component profile. The best fit is overlaid in red, and the fit residuals (data-model) are shown on top of each frame. The methanol form, quantum numbers ($J_K$), and rest frequency (GHz) are given in the bottom right corner for each line. For B4 lines, the two visits are shown separately, with their respective fits. The ALMA band is given in the lower left corners.}
\label{fig:BestFitLineProfile}
\end{center} \end{figure*}

The best-fit parameters for the above model are given in Table~\ref{tab:BestFitLineProfile}, while the fits and residuals are shown in Fig.\,\ref{fig:BestFitLineProfile}. The latter figure shows that all the lines are remarkably well reproduced with this minimal two-Gaussian model (reduced $\chi^2$ of 2.0 for 50 free parameters), with residuals below 1\% of the continuum level and without any obvious features broader than a few \kms. We therefore refrained from invoking a higher degree of complexity for the model by adding more velocity components. 

The methanol covering factors are in the $0.6$--$0.8$ range, that is, smaller than \fc\ previously measured toward \PKS1830\ for any other species (e.g., \citealt{mul14a}), suggesting that methanol has a more compact distribution than other molecules. The change in \fc\ between the two B4 tunings, at the start and end of our campaign, is $\sim 7$\%, which could indicate small morphological changes in \PKS1830. We find that, at a given time, \fc\ is always increasing with frequency (i.e., \fc(B4a) versus \fc(B6), on one hand, and \fc(B5) versus \fc(B7), on the other hand). This could be interpreted as evidence of a smaller continuum size at higher frequency, for example due to a propagation effect in the ISM of the intervening galaxy (see, e.g., \citealt{gui99,mar13}). We refer the reader to Sect.\,\ref{sec:chromaticQSR} for a more detailed discussion of the covering factor and the interpretation of its frequency dependence. We also discuss the robustness of our \fc\ fit measurements further in Appendix\,\ref{app:fc}.

\begin{table*}[ht!]
\caption{Best-fit parameters for the common methanol absorption profile (see Eq.\,(\ref{eq:commonprofile})).}
\label{tab:BestFitLineProfile}
%\tiny
\begin{center} \begin{tabular}{lccccc}
\hline \hline
\multicolumn{6}{c}{\em Velocity profile: Two-Gaussian velocity components} \\
\hline
& $G_k$ & $v_i$ & $\Delta v_k$ & $(\int \tau dv)_k$ & \\
&       & (\kms) & (\kms) & (\kms) & \\
\hline
& $G_1$ & $v_i$ (see below) & $5.36 \pm 0.02$ & $6.60 \pm 0.15$ & \\
& $G_2$ & $v_i + (3.08 \pm 0.04)$ & $14.5 \pm 0.06$ & $2.71 \pm 0.06$ & \\
\hline
%\multicolumn{5}{c}{\em Line parameters and \fc} \\
\hline
\multicolumn{6}{c}{\em Line and tuning parameters} \\
\hline
 & Rest freq.  & \fc & $\gamma_i$ & $v_i$ & $\delta v_{\nu}$ $^{(a)}$ \\
       & (GHz)      &     &            & (\kms) & (\kms)\\
\hline

B4a & 302.370 & $0.567 \pm 0.007$ & $0.226 \pm 0.005$ & $-4.575 \pm 0.029$ & 0.022\\
& 303.367 &  & $1.017 \pm 0.023$ & $-4.708 \pm 0.011$ & 0.020\\
& 304.208 &  & $0.820 \pm 0.018$ & $-4.674 \pm 0.012$ & 0.020\\
& 305.474 &  & $0.438 \pm 0.009$ & $-4.598 \pm 0.017$ & 0.019\\
& 303.693 $^{(b)}$ &  & $(164 \pm 7) \times 10^{-4}$ & $-4.606 \pm 0.335$ & 0.093\\
& 304.494 $^{(b)}$ &  & $(133 \pm 6) \times 10^{-4}$ & $-4.684 \pm 0.414$ & 0.093\\

B4b & 302.370 & $0.611 \pm 0.006$ & $0.238 \pm 0.005$ & $-4.852 \pm 0.018$ & 0.022\\
& 303.367 &  & $1.045 \pm 0.021$ & $-5.018 \pm 0.007$ & 0.020\\
& 304.208 &  & $0.852 \pm 0.017$ & $-4.967 \pm 0.008$ & 0.020\\
& 305.474 &  & $0.457 \pm 0.009$ & $-4.889 \pm 0.011$ & 0.019\\
& 303.693 $^{(b)}$ &  & $(169 \pm 7) \times 10^{-4}$ & $-4.972 \pm 0.216$ & 0.093\\
& 304.494 $^{(b)}$ &  & $(138 \pm 6) \times 10^{-4}$ & $-5.033 \pm 0.264$ & 0.093\\

B5 & 350.905 $^{(c)}$ & $0.675 \pm 0.009$ & $1$ & $-4.868 \pm 0.010$ & 0.018\\
& 350.688 &  & $0.150 \pm 0.002$ & $-5.025 \pm 0.039$ & 0.021\\

B6 & 423.440 & $0.639 \pm 0.015$ & $0.318 \pm 0.009$ & $-4.661 \pm 0.014$ & 0.016\\
& 423.469 &  & $0.206 \pm 0.006$ & $-4.621 \pm 0.020$ & 0.015\\
& 423.538 &  & $0.076 \pm 0.002$ & $-4.554 \pm 0.049$ & 0.015\\
& 423.675 &  & $0.018 \pm 0.001$ & $-4.519 \pm 0.192$ & 0.015\\
& 423.913 &  & $0.006 \pm 0.001$ & $-4.371 \pm 0.593$ & 0.013\\
& 445.571 &  & $0.616 \pm 0.019$ & $-4.661 \pm 0.009$ & 0.014\\
& 447.118 &  & $0.011 \pm 0.001$ & $-4.540 \pm 0.316$ & 0.015\\

B7 & 520.179 & $0.754 \pm 0.008$ & $0.882 \pm 0.018$ & $-4.850 \pm 0.006$ & 0.013\\

\hline
\end{tabular} \end{center}
\tablefoot{$(a)$ Velocity uncertainty corresponding to that of the line rest frequency listed in the CDMS, given in Table~\ref{tab:line-methanol}; 
$(b)$ line of $^{13}$CH$_3$OH. The opacity scaling factor was tied to that of the corresponding $^{12}$CH$_3$OH transition with a $^{12}$CH$_3$OH/$^{13}$CH$_3$OH ratio, which was fit with a value $62 \pm 3$.
$(c)$ Line taken as reference for the integrated opacity.}
\end{table*}

\section{Excitation of methanol lines} \label{sec:excitation}

The physical conditions along the southwest line of sight to \PKS1830\ have been investigated through multi-transition excitation studies using different species \citep{hen08,mul13}, finding \nH2\ of the order of a few $10^3$~\ccm\ and $\Tkin \sim 50-80$~K, assuming the same physical conditions for the entire absorption profile. Under these conditions, the excitation of molecules with a high electric dipole moment ($> 1$~Debye) remains mostly coupled to the photons from the cosmic microwave background ($\Tcmb = 5.14$~K at $z=0.89$)\footnote{The electric dipole moment of methanol is $\mu_a = 0.885$~D for $a-$type transitions (with $\Delta K =0$) and $\mu_a = 1.44$~D for $b-$type transitions $(\Delta K =\pm1)$ \citep{sas81}.}. Nevertheless, there is growing evidence for chemical segregation between the different velocity components (e.g., \citealt{mul11,mul16b,mul20a}), potentially connected to different temperature and density conditions. In particular, the detection of deuterated species with large deuterium fractionation for the $v=-5$~\kms\ velocity component suggests a relatively cold gas temperature, $<30$~K \citep{mul20a}, in contrast to the \Tkin\ values given above.

In order to investigate the excitation of our methanol lines, we used the non-LTE radiative transfer code RADEX \citep{vdtak07} with collisional rate coefficients from \cite{pot04}. These collisional rate coefficients were calculated for temperatures between 5~K and 200~K and rotational states up to $J=9$. More recently, \cite{rab10} extended the rate calculations up to the level $J=15$, although only for temperatures starting at 10~K. Para-H$_2$ is assumed to be the main collisional partner, since it is expected to be the dominant form at low temperatures ($T < 100$~K). The background temperature was fixed to $\Tcmb = 5.14$~K, and we did not consider additional heating, either by the ambient radiative field or cosmic rays. We adopted a uniform sphere geometry and took the two-Gaussian line profile (i.e., fixing their velocity centroids, line widths, and integrated opacity ratios) and the covering factors determined in Table~\ref{tab:BestFitLineProfile}. We explored a density range between $10^3$ and $10^6$~\ccm\ (in logarithmic steps) and kinetic temperatures between 6 and 100~K, varying also the $G_1$ column density (that of $G_2$ is tied by their integrated opacity ratio) and the A/E methanol abundance ratio. We note that the two velocity components are treated as if they were independent, given they are mostly kinematically decoupled, that is, radiative transfer issues were neglected. For each model, we constructed the RADEX methanol spectrum and calculated the resulting $\chi^2$, summing over the different tunings $i$ as
\begin{equation}
\chi^2 = \sum_i \left (\frac{S_{\rm ALMA}^i - S_{\rm RADEX}^i}{\sigma_i} \right )^2,
\end{equation}
where $\sigma_i$ is the rms noise of the observations in tuning $i$. For this exercise, the two visits of the B4 tuning were averaged, and we took their average \fc. The $^{13}$CH$_3$OH transitions were also modeled, scaling the opacity of their corresponding $^{12}$CH$_3$OH transition by a ratio of 62 (see Sect.\,\ref{sec:C1213}), although they were not taken into account in the $\chi^2$ minimization.

\begin{figure*}[h] \begin{center}
\includegraphics[width=\textwidth]{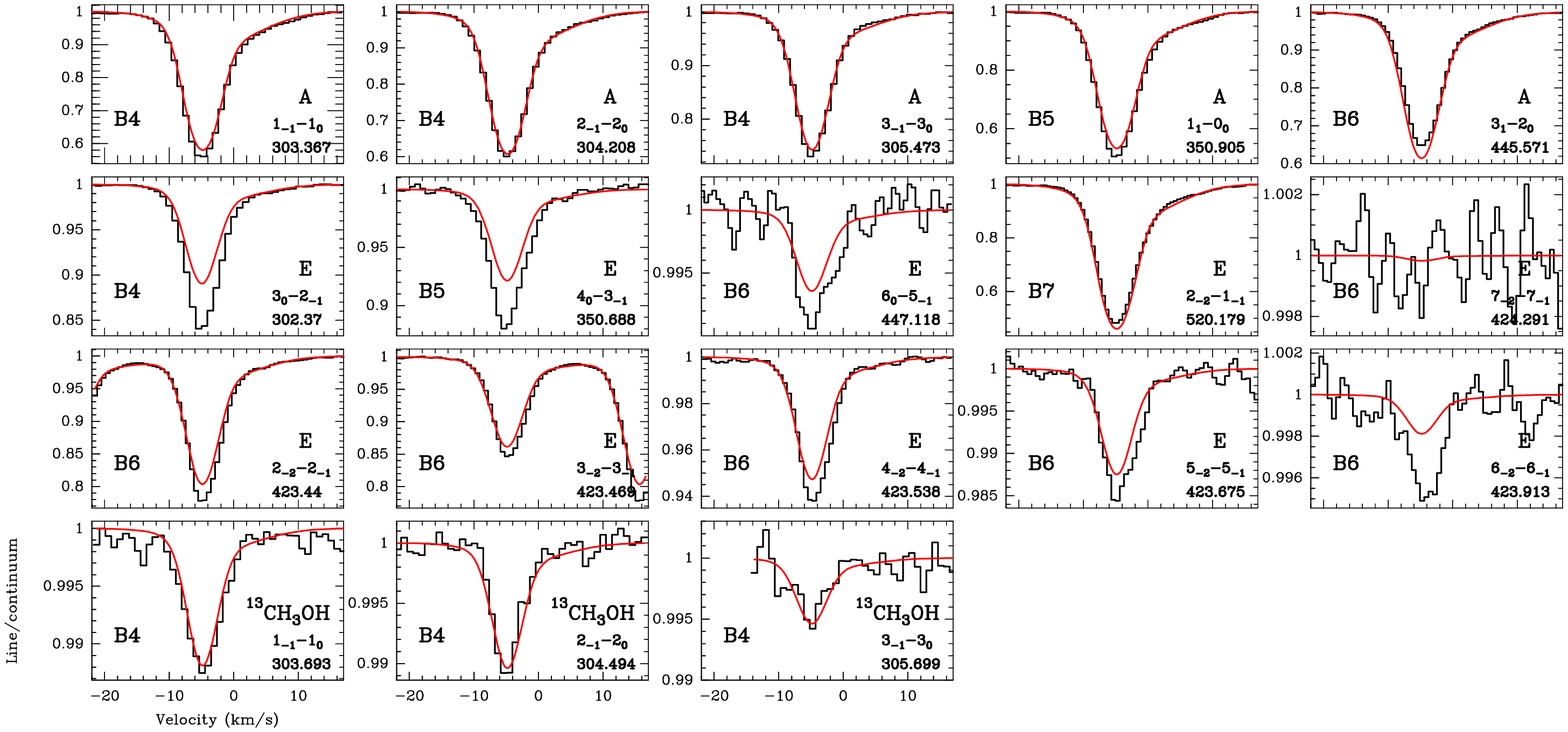}
\caption{RADEX excitation best-fit model ($\Tkin=9$~K, $\nH2= 1.3 \times 10^5$~\ccm, A/E=1, and $\Ncol(A)=1.8 \times 10^{14}$~\cmsq\ and $0.7 \times 10^{14}$~\cmsq\ for the $G_1$ and $G_2$ velocity components in Table~\ref{tab:BestFitLineProfile}, respectively), overlaid in red on top of the ALMA methanol spectra toward the southwest image of \PKS1830. The methanol type, quantum numbers ($J_K$), and rest frequency (GHz) are given in the bottom right corner for each transition. The ALMA band is given in the bottom left of each frame. The last row shows the $^{13}$CH$_3$OH  spectra (all of A type), with the RADEX fit of the corresponding transitions of the main isotopologue, scaled down by a factor 62; hence, these lines were not taken into account in the $\chi^2$minimization.}
\label{fig:fig-spec-RADEX}
\end{center} \end{figure*}

\begin{figure}[h] \begin{center}
\includegraphics[width=8.8cm]{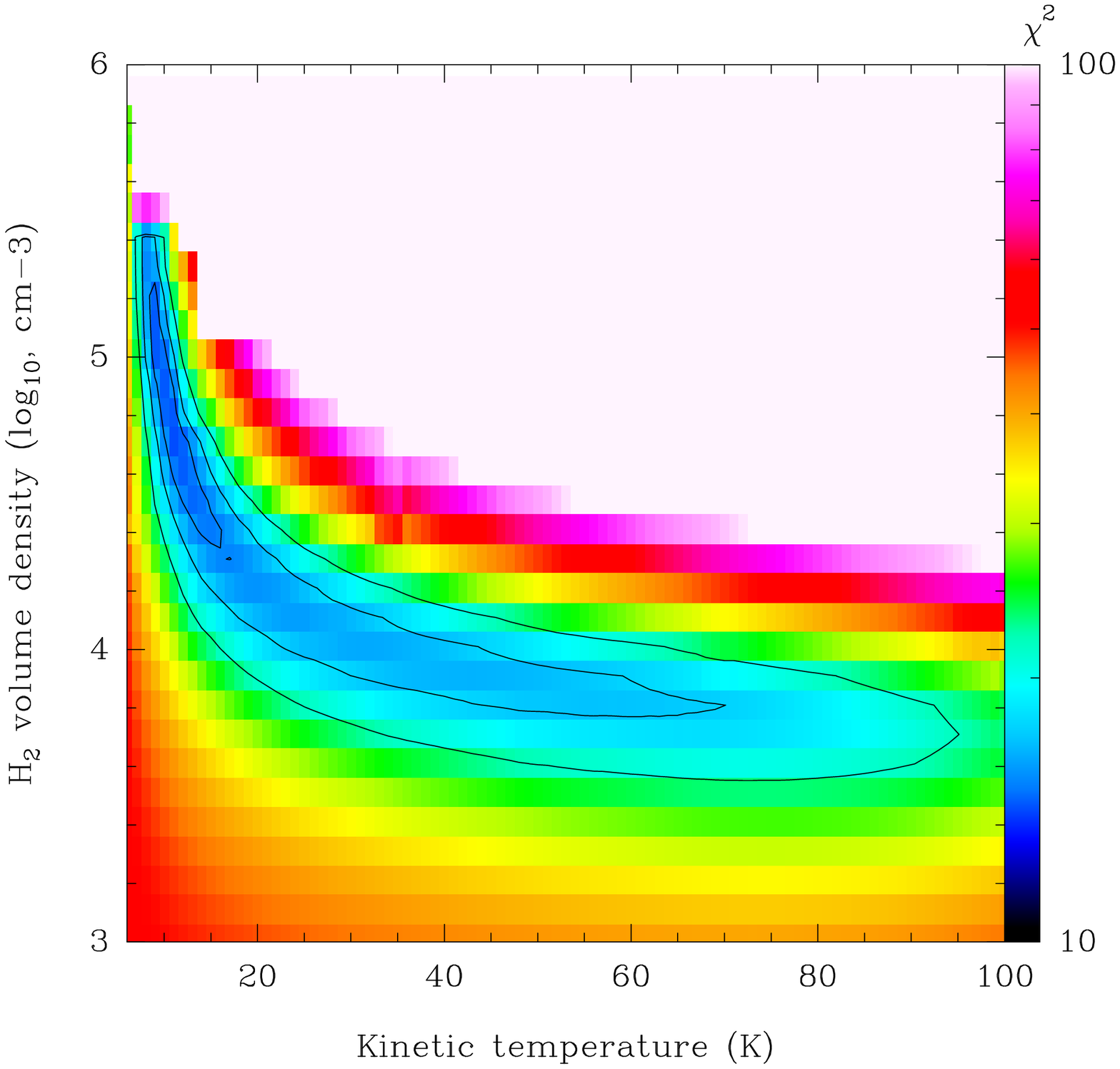}
\caption{Reduced $\chi^2$ results for the RADEX \Tkin-\nH2\ grid search. The black contours are drawn at $\chi^2_{min}+2.3$,
$\chi^2_{min}+4.6$, and $\chi^2_{min}+9.2$, corresponding to $1\sigma$, $2\sigma$, and $3\sigma$ confidence levels, respectively. The column densities of A-CH$_3$OH were set to $1.8 \times 10^{14}$~\cmsq\ and $0.7 \times 10^{14}$~\cmsq\ for the $G1$ and $G2$ velocity components, respectively, and the A/E abundance ratio was set to unity. Kinetic temperatures were explored between 6 and 100~K, in steps of 1~K, and the H$_2$ densities were set in \ccm-logarithmic values between 3 and 6, in steps of 0.1.}
\label{fig:RADEXgrid}
\end{center} \end{figure}

The best-fit solution ($\chi^2_r = 13$) is found for column densities of A-CH$_3$OH of $1.8 \times 10^{14}$~\cmsq\ and $0.7 \times 10^{14}$~\cmsq\ for $G_1$ and $G_2$, respectively, an A/E abundance ratio of 1, $\Tkin=9$~K, and $\nH2 = 1.3 \times 10^5$~\ccm. The corresponding RADEX model spectra are overlaid on the ALMA spectra in Fig.\,\ref{fig:fig-spec-RADEX}. It is not straightforward to define uncertainties in the four-dimensional parameter space. Nevertheless, in Fig.\,\ref{fig:RADEXgrid} we plot the reduced $\chi^2$ values obtained from the fit as a function of \Tkin\ and \nH2, after fixing the column densities of $G_1$ and $G_2$ to the above values, and A/E to unity. The equivalent $1\sigma$ contour, defined at a value Min($\chi^2_r)+2.30$, indicates a region with \Tkin\ between $8 - 17$~K and \nH2\ between $2 \times 10^4 - 2 \times 10^5$~\ccm, meaning a cold and dense methanol gas, although a solution with a higher temperature ($\Tkin >20$~K) and lower density ($\nH2 \lesssim 10^4$~\ccm) is not completely excluded. We note, a posteriori, that the Rayleigh-Jeans approximation ($h\nu << k_BT$) is not applicable for temperatures $T \lesssim 15-25$~K for the submillimeter lines targeted here, preventing the use of the traditional rotation diagram analysis.

The match between the observed and RADEX spectra is remarkably good for transitions of A-CH$_3$OH (Fig.\,\ref{fig:fig-spec-RADEX}, upper row). However, for most lines of E-CH$_3$OH (except the ground state transition), the RADEX spectra underestimate the absorption depths. Removing the two lines with the largest mismatch from the $\chi^2_r$ calculation, namely the $3_0-2_{-1}$ and $4_0-3_{-1}$~E lines, we obtained an improved value $\chi^2_r=9$. This is a similar situation to that encountered by \cite{dap17}, who failed to reproduce the intensity of one of their E-CH$_3$OH lines in the dark cloud L1498, in spite of its very simple cloud structure. We note that \cite{dap17} used methanol collisional rates from \cite{rab10}. We also explored solutions based on these rates, and found only minor improvements for few E-CH$_3$OH lines, except the two lines mentioned above, which are still not well reproduced. On the other hand, we found some inconsistencies in the $A_{ul}$ Einstein coefficients in the LAMDA file of A-CH$_3$OH associated with the \cite{rab10} rates, and for this reason, we continued to use the \cite{pot04} rates for both A- and E-CH$_3$OH.

We do not have a clear explanation for the above mismatch seen in a few lines. Possible paths for further investigations could be, for example, allowing different conditions between the A- and E-types, relaxing the single \Tkin/\nH2-set model to account for a multiphase medium, allowing different physical conditions and \fc\ for the different velocity components, and considering collisions with ortho-H$_2$ \citep{rab10}. All these would require more free parameters to fit, adding significant complexity to the problem, as well as additional inputs (the ortho-to-para ratio of H$_2$, calculations of the corresponding rates, etc).

Finally, we note that according to our RADEX simulations, the excitation temperature of some lines of E-CH$_3$OH are below \Tcmb. This does not impact the spectra simulated by RADEX, but it is an interesting curiosity, and we further discuss this anti-inversion effect in Appendix~\ref{sec:AntiInversion}.

\subsection{The A/E ratio}

From the results of our RADEX grid search, we estimate an abundance ratio A/E~$= 1.0 \pm 0.1$. This ratio corresponds to a nuclear spin temperature $\gtrsim 20$~K (see, e.g., Fig.\,7 by \citealt{wir11}). Since there are no allowed radiative or collisional transitions between A- and E-CH$_3$OH species, the spin temperature should reflect the initial temperature at the formation stage of the molecule, unless other mechanisms in the ISM drive the population distribution toward the equilibrium ratio at high temperatures, A/E = 1 (see, e.g., \citealt{fri88}). Observationally, the case is not clear (e.g., \citealt{wir11}), and it is difficult to establish the processes of methanol formation based on the A/E ratio alone.

\subsection{Methanol relative abundance} \label{sec:abundance}

By adding the column densities of A- and E-CH$_3$OH over the two velocity components $G_1$ and $G_2$ , we estimate a total methanol column density of $\sim 5 \times 10^{14}$~\cmsq\ along the southwest line of sight. We further take the H$_2$ column density of $2 \times 10^{22}$~\cmsq, estimated by \cite{mul14a} using CH spectra (obtained in 2012), as an H$_2$ proxy. This yields an average methanol abundance relative to H$_2$ of CH$_3$OH/H$_2 \sim 2 \times 10^{-8}$ toward the southwest image of \PKS1830. We note that this is a rough estimate, because there were some variations in the line profile between 2012 and 2019, and because the two velocity components $G_1$ and $G_2$ could have different molecular compositions. Nevertheless, this abundance falls between methanol abundances in Galactic cold dark clouds ($\sim 10^{-9}-10^{-8}$) and hot cores ($\gtrsim 10^{-7}$). We find a difference of at least two orders of magnitude for the CH$_3$OH/H$_2$ abundance ratio between the southwest and northeast lines of sight, clearly denoting their different chemical compositions (e.g., \citealt{mul17}, their Table~4).

\subsection{The $^{12}$CH$_3$OH/$^{13}$CH$_3$OH ratio} \label{sec:C1213}

We detected absorption from three transitions of $^{13}$CH$_3$OH (Fig.\,\ref{fig:spec}d), which are the counterparts of the three $^{12}$CH$_3$OH transitions covered in the same B4 tuning. In the construction of the global line profile of the methanol absorption (Sect.\,\ref{sec:GlobalProfile}), we tied the opacities of the $^{13}$CH$_3$OH lines to those corresponding to the $^{12}$C- main isotopologue, considering only the two brightest ones, since the third one has a low S/N and falls close to the edge of the spectral window (see Fig.\,\ref{fig:spec}d). Accordingly, we determined a $^{12}$CH$_3$OH/$^{13}$CH$_3$OH ratio of $62 \pm 3$. This value falls between the $^{12}$C/$^{13}$C-isotopologue ratios $\sim 30-40$, measured from HCO$^+$, HCN, and HNC \citep{mul06,mul11}, and $97 \pm 6$ determined for CH$^+$ \citep{mul17}, respectively. Therefore, this result suggests that fractionation effects play an important role and that a more detailed analysis (beyond the scope of this paper) will be necessary to derive the elemental $^{12}$C/$^{13}$C ratio.

\subsection{The nature of the $v=-5$~\kms\ velocity component}

With the new estimate of physical conditions from methanol excitation, there is growing evidence that the $v=-5$~\kms\ velocity component is associated with the analog of a Galactic cold dark cloud.
\cite{mul20a} gave a list of arguments to this effect:
{\em (i)} the strong deuterium fractionation of ND, NH$_2$D, and HDO (up to 100 times larger than the primordial D/H ratio), pointing toward a temperature $\lesssim 30$~K, which is consistent with the cool and dense solution determined above;
{\em (ii)} the chemical composition, with relative enhancement of molecules typically associated with dense cores (e.g., CH$_3$OH, HC$_3$N, N$_2$H$^+$, SO$_2$);
{\em (iii)} the fast time variability suggesting a typical size of the order of 1~pc \citep{mul08, mul20a};
{\em (iv)} hints at temperature-velocity gradients suggesting the existence of substructures in the absorbing gas, which adds to the evidence of a multiphase absorbing gas column \citep{sch15,mul16a,mul16b,mul17}.

We note that, from their Very Long Baseline Interferometry (VLBI) imaging of the 12.2~GHz transition of methanol, \cite{mar17} found a possible offset between the strongest absorption and the peak of the continuum emission, suggesting that the size of the absorbing material is in the range 1--10 pc. Therefore, there is hope that a direct measurement of the size of the methanol absorbing cloud could be possible with future observations.

It is interesting to compare our study of methanol absorption in the $z=0.89$ absorber with that of methanol emission from the L1498 dark cloud in the Milky Way Taurus-Auriga complex by \cite{dap17}. They found comparable excitation conditions: $\Tkin = 6 \pm 1$~K and $\nH2 \sim 3 \times 10^5$~\ccm, and A/E = $1.00 \pm 0.15$. On the other hand, the FWHM of the methanol emission in L1498 is much narrower, $< 0.2$~\kms, suggesting a low degree of turbulence. At $z=0.89$, the heating by CMB photons brings an additional 2.4~K and it is possible that the heating by cosmic rays is higher than in the surroundings of L1498. Indeed, \cite{mul16a} estimated a cosmic-ray ionization rate of atomic hydrogen $\zeta_H \sim 3 \times 10^{-15}$~s$^{-1}$ along the southwest line of sight toward \PKS1830, which is slightly higher than in the Milky Way at a comparable galactocentric radius.

\section{Invariance of $\mu$ and systematics} \label{sec:mu}

\begin{figure*}[h] \begin{center}
\includegraphics[width=\textwidth]{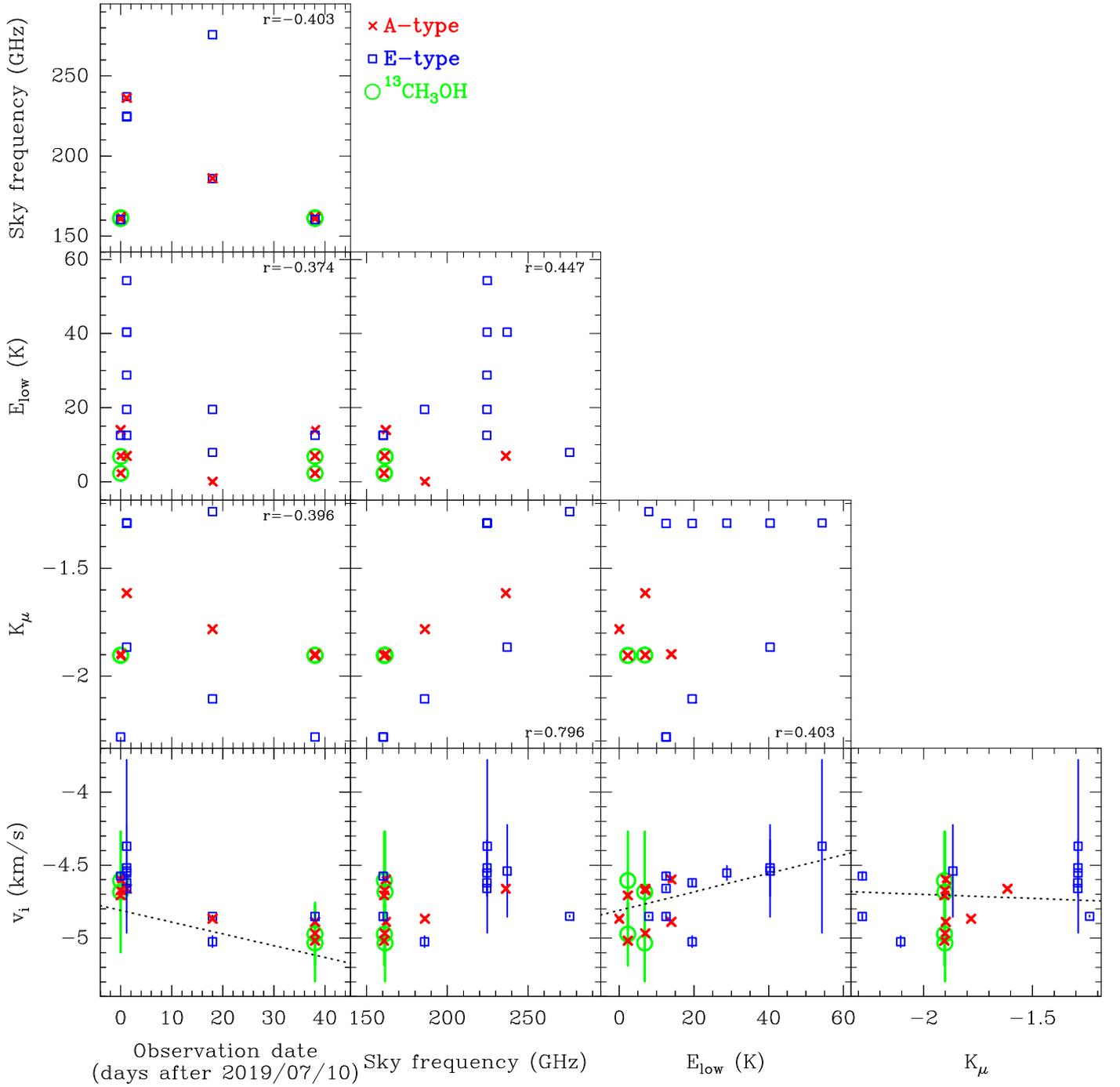}
\caption{Distribution matrix of relative bulk velocity offsets ($\delta v$), low-level energies ($E_{low}$), proton-electron sensitivity coefficients (\Kmu), sky frequencies, and observation epochs, for observed methanol lines (Tables\,\ref{tab:BestFitLineProfile} and \ref{tab:line-methanol}). The Pearson $r$ coefficients of the correlation matrix are given between the different variables. The data points corresponding to A-CH$_3$OH and E-CH$_3$OH transitions are marked with red crosses and blue squares, respectively. Those of $^{13}$CH$_3$OH are indicated with green circles. The best results of the multivariate linear regression fit between the $v_i$, time, \Elow, and \Kmu\ (see Table~\ref{tab:MultiVariateFit}) are shown in the corresponding panels.}
\label{fig:linedistribution}
\end{center} \end{figure*}

\begin{table*}[ht!]
\caption{Best-fit parameters for the multivariate regression analysis of the $v_i$ measurements.}
\label{tab:MultiVariateFit}
\begin{center} %\tiny
\begin{tabular}{cccccccc}
\hline \hline
Variable & Multivariate regression & \Dmu &  $c \times \Gamma_t$ & $c \times \Gamma_E$ & $c \times \Gamma_\nu$ & $c \times \Gamma_0$ & $\chi^2_r$ \\
& $v_i / c = $ & ($10^{-7}$) & (m\,s$^{-1}$\,d$^{-1}$) & (m\,s$^{-1}$\,K$^{-1}$) & (m\,s$^{-1}$\,GHz$^{-1}$) & (\kms) & \\
\hline
$\Kmu$ & $\Dmu \cdot \Kmu + \Gamma_0$ & $2.9 \pm 3.1$ & -- & -- & -- & $-4.64 \pm 0.16$ & 27 \\
$t$   & $\Gamma_t \cdot t + \Gamma_0$ & -- & $-8.1 \pm 0.9$ & -- & -- & $-4.66 \pm 0.02$ & 5.3 \\
$\Elow$   & $\Gamma_E \cdot \Elow + \Gamma_0$ & --  & -- & $9.3 \pm 4.9$ & -- & $-4.87 \pm 0.05$ & 23.8 \\
$\nu$ & $\Gamma_\nu\cdot\nu + \Gamma_0$ & -- & -- & -- & $2.5 \pm 3.7$  & $-4.88 \pm 0.14$ & 27.4 \\
$\Kmu, t$ & $\Dmu\cdot\Kmu + \Gamma_t\cdot t + \Gamma_0$ & $-1.3 \pm 1.5$  & $-8.4 \pm 0.9$ & -- & -- & $-4.72 \pm 0.07$ & 5.3 \\
$\Kmu, \Elow$ & $\Dmu\cdot\Kmu + \Gamma_E\cdot\Elow + \Gamma_0$ & $2.0 \pm 3.1$ & -- & $8.8 \pm 5.1$ & -- & $-4.76 \pm 0.17$ & 24.5 \\
$\nu, t$ & $\Gamma_\nu\cdot\nu  + \Gamma_t\cdot t + \Gamma_0$ & -- & $-8.4 \pm 0.9$ & -- & $-1.8 \pm 1.7$  & $-4.59 \pm 0.07$ & 5.2 \\
$\Kmu, t, \Elow$ & $\Dmu\cdot\Kmu  + \Gamma_t\cdot t + \Gamma_E\cdot\Elow + \Gamma_0$ & $-1.8 \pm 1.2$ & $-8.1 \pm 0.8$ & $6.3 \pm 1.9$ & -- & $-4.81 \pm 0.06$ & 3.5 \\
\hline
\end{tabular} \end{center}
%\tablefoot{}
\end{table*}

The primary goal of our ALMA methanol campaign was to test the possibility of cosmological evolution in the proton-electron mass ratio $\mu$ from $z=0.89$ to today. For this, we targeted most of the strongest submillimeter methanol transitions, for which a high spectral resolution ($\nu/\delta \nu \sim 3 \times 10^5$) and a high S/N ($\gtrsim 500$) could be obtained with the ALMA observations. Although the spread in \Kmu\ is limited, $\Delta\Kmu \sim 1$ \footnote{The spread in \Kmu\ with methanol transitions detected at centimeter wavelengths (e.g., by \citealt{ell12,bag13a,bag13b,kan15}) can reach values of up to $\sim 30$ (see \citealt{jan11a,jan11b, lev11}).},
the high S/N, the short time span of the observations (all performed within $<40$~days), and the limited frequency range ($\Delta \nu / \nu \sim 0.6$) make it possible to obtain a reliable constraint, addressing systematic issues. Such systematic effects arise from two aspects, the absorption lines themselves and the background source structure. The former includes temporal variations of the absorption profile \citep{mul08,mul14a} and chemical or excitation segregation (gas covering factor, co-spatiality of species and lines with different excitation). The latter stems from the chromatic dependence of the background source morphology; for example, the effect of the Einstein ring at low radio frequencies \citep{kan15,com21}, frequency-dependent size of the continuum emission \citep{gui99}, and the core shift due to opacity effects along the jet \citep{mar13}, which can result in a changing line of sight or covering factor at different frequencies. Some assessments of the above systematics have been discussed in earlier studies of $\mu$-invariance using \PKS1830\ \citep{bag13b,kan15}.

We note that the uncertainties on our $v_i$ measurements are, in some cases, of the same order as the methanol rest frequency uncertainties listed in the CDMS database. Those entries are the results of a global fit to the CH$_3$OH rotational spectrum \citep{xu08}, based on a total of 24600 measured frequencies. While adequate for most astronomy studies, they do not have the best possible accuracy that could be achieved in the lab. Actually, the post-fit error given in the CDMS ($\sim 10$~kHz for the submillimeter CH$_3$OH lines considered here) is much smaller than the error of an individual measurement (i.e., 50~kHz, \citealt{xu08}). For $^{13}$CH$_3$OH, the CDMS entries are based on the work of \cite{xu97}, with rest frequency uncertainties of 50~kHz. In order to take the rest frequency uncertainties into account, we thus added them quadratically to the uncertainties of our $v_i$ measurements.

On the other hand, the hyperfine structure splitting of methanol lines (of the order of 10~kHz; see, e.g., \citealt{cou15,lan16}) is negligible compared to the width of methanol absorption toward \PKS1830\ and also remains small compared to the uncertainties of our $v_i$ measurements (see also the discussion of methanol hyperfine structure and its relevance to $\mu$-invariance determination by \citealt{dap17}). Besides, \cite{lan18} noticed that the methanol hyperfine structure could be impacted by the Zeeman effect in case of strong magnetic fields ($> 30$~mG, as they quote). Again, this effect is completely negligible, given the line widths of the absorption toward \PKS1830, the uncertainties in our $v_i$ fit measurements, and the expected magnetic fields in molecular clouds.

Following Eq.\,(\ref{eq:mu}) and taking the bulk velocities $v_i$ in Table~\ref{tab:BestFitLineProfile} at face value (including transitions of $^{13}$CH$_3$OH), weighted by the inverse of their squared uncertainties (to which we quadratically added the rest frequency uncertainties), together with the corresponding \Kmu\ (see Table~\ref{tab:line-methanol} and Appendix~\ref{sec:Kmu}), we obtained $v_i/c = (2.9 \pm 3.1) \times 10^{-7} \times \Kmu + (-4.64 \pm 0.16)/c$, with errors quoted at a 1$\sigma$ confidence level (CL). This provides us with a first upper limit $| \Dmu | < 9.4 \times 10^{-7}$ at 3$\sigma$ CL. However, the reduced chi-squared is high ($\chi^2_r = 27$), and we clearly see a systematic drift of $\sim 0.3$~\kms\ between B4a and B4b visits (Fig.\,\ref{fig:linedistribution}, lower left panel), as well as an apparent correlation between $v_i$ and the low-level energy (Fig.\,\ref{fig:linedistribution}, lower row, third panel) which was already noticed by \cite{bag13b}. Therefore, we further performed several multivariate regression analyses, taking the different systematic effects into account to improve the statistical description of our data.

The different possible variables to check for systematic effects are the observing epoch, the frequency, and the lower level energy of the transitions. Their correlation matrix is shown in Fig.\,\ref{fig:linedistribution}. We did not consider the frequency and \Kmu\ as simultaneous variables, because there is a clear correlation between them, with a Pearson coefficient $r\sim0.8$  (see also \citealt{jan11a}). All other pairs of variables do not show significant correlation, with $|r| \lesssim 0.4$. The results of our different trials are given in Table~\ref{tab:MultiVariateFit}, with different variable combinations for a multivariate linear regression analysis of the following type:
\begin{equation} \label{eq:MultivariateLinearRegression}
  v_i/c = (\Dmu \cdot K_{\mu}) + (\Gamma_{t} \cdot t) + (\Gamma_E \cdot E_{low}) + \Gamma_0,
\end{equation}
where $\Gamma_t$ and $\Gamma_E$ are the coefficients of the correlation with time and \Elow, respectively, and $\Gamma_0$, a constant absorbing the velocity reference. Thus, Eq.\,(\ref{eq:MultivariateLinearRegression}) is a modified version of Eq.~(\ref{eq:mu}), with time and \Elow\ systematics included.

Our first obvious result is that the fit quality drastically improves when a linear dependence with observing epoch is allowed, with $\chi^2_r$ values decreasing from $>24$ down to $< 6$. The correlation between $v_i$ and time is strong, with a coefficient $| \Gamma_t | \sim 8$~m\,s$^{-1}$\,d$^{-1}$ determined at $\sim 10\sigma$. We have further tested different time functions, such as quadratic polynomial and step functions, although the $\chi^2_r$ did not decrease drastically ($\chi^2_r \sim 3-4$), while adding one more free parameter to fit when we only have three separated observing epochs. Thus we restrained ourselves to linear regression.

Besides the observing epoch, we found a weaker correlation between the $v_i$ and \Elow, with a coefficient $\Gamma_E$ determined at $3\sigma$ at best. Replacing the \Kmu\ by frequencies in Eq.\,(\ref{eq:MultivariateLinearRegression}), we did not find a significant correlation between $v_i$ and frequencies, showing that the chromatic effects from the background blazar have little impact on our constraint on \Dmu. Therefore, temporal variations of the absorption profile are the dominant systematic effect in our data. 

Overall, the uncertainties on \Dmu\ do not change drastically throughout the different regressions, varying within the range $(1.2-1.5) \times 10^{-7}$ at $1\sigma$ CL once the $v_i$ shift with time is taken into account. The final multivariate regression with \Kmu, observing epoch, and \Elow\ yields the lowest $\chi^2_r$ value ($\chi^2_r =3.5$) and gives the constraint $\Dmu = (-1.8 \pm 1.2) \times 10^{-7}$ ($1\sigma$ CL). Assuming a Gaussian distribution of errors, we take the 1-$\sigma$ uncertainty\footnote{We recall that the fit uncertainties were obtained as the square root of the diagonal elements of the covariance matrix multiplied by the reduced chi squared.} of this fit result to obtain an upper limit of $3.6 \times 10^{-7}$ at $3\sigma$ (99.7\% CL) for $| \Dmu |$ at a look-back time of half the present age of the Universe. Our results expand the study of \cite{bag13b} with a robust handling of the systematics, which we discuss in more detail in the following subsections.

It would be tempting to combine our ALMA observations of submillimeter methanol lines with other lines providing a large spread in $\Delta \Kmu$, notably the 12.2~GHz transition which has a $\Kmu = -32$, thus offering a large lever arm for constraints on \Dmu. This line was observed by several groups (\citealt{ell12, bag13a, bag13b, kan15, mar17}) between 2011 and 2013. The line centroid velocity is found to vary between $-5$~\kms\ and 0~\kms\ (with typical uncertainties of $\sim 1$~\kms) in fits with a single-Gaussian velocity component. The FWHM is found in the 8--20~\kms\ range. These differences may be due to temporal variations of the absorption profile (even if those are expected to be less than at millimeter wavelengths, \citealt{all17,com21}) or to the limited S/N of some of the observations. Furthermore, \cite{kan15}, with the best S/N achieved on the 12.2~GHz line so far, made a detailed analysis of the 12.2, 48, and 60.5~GHz lines observed with the VLA between 2012 July and August. They showed that the 12.2~GHz line had a significantly different profile with respect to the two others, suggesting that it probes a different sightline. Hence, we remained conservative in our upper limit on \Dmu, and chose not to combine our results with previously published low-frequency methanol data.

\subsection{Temporal variations of the absorption profile}

The origin of the temporal variations of the $v_i$, and of the absorption profile, is not clearly established, although it is most likely due to structural changes in the morphology of the quasar, possibly with the recurrent ejection of plasmons in the jet \citep{nai05,mul08,mar13,sch15}. The geometry of the system is indeed particularly favorable for such effects with the stretching due to a magnification factor of $\sim 3$ (e.g., \citealt{mul20b}) and the blazar's jet almost exactly oriented in our direction. In fact, evidence for superluminal motions was observed by \cite{jin03} during a monitoring of the relative separation between the two lensed images of \PKS1830 with VLBI. They found jumps in the plane of the sky of up to $\sim 200$~$\mu$as within eight~months, which would correspond to apparent motions at $v \sim 8c$. Converted into the plane of the $z=0.89$ absorber, such a motion would cover a projected distance of $\sim 1$~pc in five months.

Within the same time interval of five months, the rate of change in velocity with time of $\Gamma_t = 8$~m\,s$^{-1}$\,d$^{-1}$ obtained in our regression analysis would result in a velocity drift of $\sim 1$~\kms. If the cloud responsible of the methanol absorption is roughly placed on the size-line width relationship for Galactic molecular clouds \citep{lar81}, it would have a line width of $\sim 1$~\kms\ for a size of 1~pc. Therefore, the temporal variations observed in the line profile of methanol within a couple of months are naturally explained by superluminal motions in the continuum illumination, with the recurrent ejection of plasmons in the jet.

\subsection{Temperature gradient}

The correlation between line velocity and \Elow\ was already noticed by \cite{bag13b} (with the same sign, i.e., higher velocity toward higher \Elow) and suggests a temperature gradient in the absorbing gas. Indeed, there is strong evidence of inhomogeneities along the southwest line of sight toward \PKS1830\ \citep{sat13,sch15,mul16b,mul20a}. The physical conditions were investigated by \cite{hen08} and \cite{mul13}, both finding a kinetic temperature of $\sim 50-80$~K and an H$_2$ density of $\approx few \times 10^3$~\ccm\ over the full line profile. If the two velocity components $G_1$ and $G_2$ have different physical properties (i.e., $G_1$ being associated with a cold dark cloud, and $G_2$ having a higher temperature), we would indeed expect a positive temperature gradient toward higher velocities.

\subsection{Chromatic structure of the quasar} \label{sec:chromaticQSR}

The morphology of \PKS1830\ is highly dependent on the observing wavelength. The emission from the Einstein ring has a steep spectral index, and while it starts to become dominant at low radio frequencies ($\lesssim 1$~GHz), it is vanishingly weak in the millimeter and submillimeter domains \citep{mul20b}. There, almost all of the continuum emission arises from the two bright and compact core components, which have a spectral index of $\alpha \sim -0.7$, consistently with optically thin synchrotron emission (Fig.\,\ref{fig:contvar}). The change in continuum illumination between centimeter and millimeter lines has a considerable effect on the resulting absorption profiles, as shown by \cite{com21}. However, the homogeneous set of millimeter-wave methanol lines is not expected to introduce a significant chromatic bias (e.g., by a differential illumination from the Einstein ring) in our study.

The sizes of the core images of \PKS1830 were measured at centimeter wavelengths, $\lambda$, with VLBI by \cite{gui99}. The angular size $\theta$ of the southwest image follows a steep $\theta_{\rm SW} \propto \lambda^2$ dependence, indicative of broadening due to interstellar scattering. In contrast, that of the northeast has a flatter dependence, $\theta_{\rm NE} \propto \lambda^{0.65}$, which \cite{gui99} attributed to different scattering properties by interstellar plasma between the two lines of sight through the disk of the intervening galaxy. Extrapolating their measurements to the millimeter domain, we would obtain a size of the order of 1 $\mu$as for the southwest image. However, it is questionable whether the frequency-dependence of the size, which is established at radio-centimeter wavelengths, holds in the (sub)millimeter range, in which broadening by interstellar scattering is expected to have a far less significant effect.

\begin{figure}[h] \begin{center}
\includegraphics[width=8.8cm]{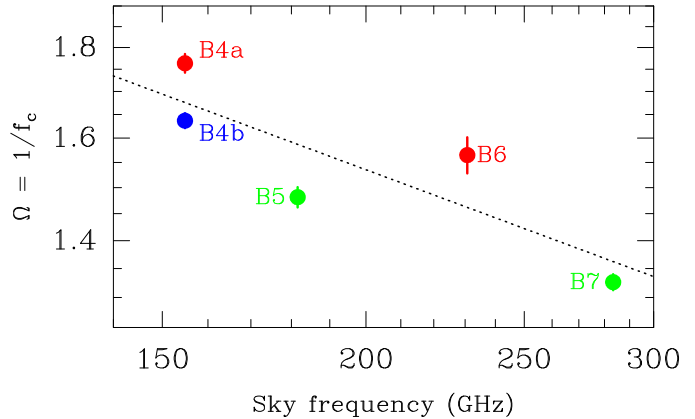}
\caption{Evolution of the inverse of the covering factor with frequency (both axes are on a logarithmic scale). The color-code indicates the observing epoch (red: 2019 Jul 10-11, green: 2019 Jul 28, and blue: 2019 Aug 17).}
\label{fig:fc-index}
\end{center} \end{figure}

If we interpret the covering factor \fc\ as the ratio between the solid angle subtended by the absorbing gas to that of the continuum emission, $\fc = \Omega_{\rm gas} / \Omega_{\rm cont}$, and if we assume that the absorbing gas has a constant solid angle, then the inverse of \fc\ gives the solid angle subtended by the continuum emission in units of the solid angle subtended by the absorbing gas. In Fig.\,\ref{fig:fc-index}, we plotted 1/\fc\ as a function of the observing frequency of our different visits. We estimate a rough relationship $1/\fc \propto \nu^{-0.3}$, where the temporal variations of the continuum intensity largely dominate the error budget on the spectral index. The slope in this figure holds if we consider either the B4a--B6 pair or the B5--B7 pair separately. We thus find a much flatter size dependence on frequency at millimeter wavelengths than in the centimeter range. As an upper limit, the size of the southwest image in the millimeter range should be of the order of a tiny fraction of a milli-arcsecond. We note that the size of the continuum emission would only change by $\sim 20$\% over the frequencies covered by our campaign. 

Finally, we discuss the impact of a core-shift effect, which denotes a change of the core peak brightness location as a function of frequency, which is driven by opacity effects (see e.g., \citealt{mar84,lob98}). The shift $\theta_{cs}$ between two frequencies $\nu_1$ and $\nu_2$ can be expressed as
\begin{equation}
\theta_{cs} = \Omega \left ( \frac{1}{\nu_1}-\frac{1}{\nu_2} \right ),
\end{equation}
where $\Omega$ is the normalized core shift.

From early ALMA observations, \cite{mar13} noticed the peculiar temporal and chromatic millimeter variability of the flux ratio between the two lensed images of \PKS1830, coincident in time with a strong gamma-ray flare. They proposed a simple model of plasmon ejection in the blazar’s jet to account for the ALMA observations and estimated a core shift of $\Omega \sim 0.8$~mas\,GHz, accounting for a magnification of three for the southwest image. Taking this value for the extreme frequencies of our ALMA campaign, we obtain a maximum core shift of $\sim 2$~$\mu$as. Projected in the plane of the $z=0.89$ absorber, this corresponds to a size of $\sim 0.02$~pc, which is much smaller than the typical size of a molecular cloud, and not sufficient to introduce a significant correlation with frequency in our regression analysis.

In summary, the frequency-dependent morphology of the blazar can introduce strong biases in studies aiming to constrain $\mu$-variations when comparing lines with a large difference in frequency (e.g., between the radio-centimeter and the millimeter and submillimeter windows). Nevertheless, all these chromatic effects remain small for the relatively narrow frequency range of our submillimeter line selection, and we did not see any significant correlation between the velocity measurements and the line frequencies.

\subsection{Is there a difference between A/E methanol types?}

A dependence of the line kinematics on the methanol type was suggested by \cite{bag13a} but later refuted by \cite{bag13b}. Taking the same combinations as in Table~\ref{tab:MultiVariateFit} but separating A- and E-CH$_3$OH in a new multivariate regression analysis, we did not see any significant differences in the resulting $\Gamma$ coefficients between the two types. The constraint on \Dmu\ is slightly deteriorated, with $1\sigma$ uncertainties $(1.5-1.8) \times 10^{-7}$, depending on the adopted regression.

% https://statsjournal.com/regression-result

\section{Summary and conclusions} \label{sec:summary}

We presented an absorption study of submillimeter lines of methanol from the $z=0.89$ molecular absorber of the lensed blazar \PKS1830, with the goals of studying methanol excitation and probing cosmological evolution in the proton-electron mass ratio $\mu$. Our main results are as follows:
\begin{itemize}
\item We detected 14 methanol lines, five from A-CH$_3$OH and nine from E-CH$_3$OH, in absorption toward the southwest image of \PKS1830, as well as three transitions of $^{13}$CH$_3$OH. None of the lines are detected in emission, implying that they are not affected by inversion, and thus do not amplify the background continuum.
\item Our excitation analysis points to a cool and dense methanol gas, with a kinetic temperature $\Tkin \sim 10-20$~K and a volume density $\nH2 \sim 10^{4-5}$~\cmsq. We estimated a methanol abundance of $\sim 10^{-8}$ relative to H$_2$. Such conditions are reminiscent of cold dark clouds in the Milky Way. 
\item We determined a methanol A/E abundance ratio of $1.0 \pm 0.1$ and a $^{12}$CH$_3$OH/$^{13}$CH$_3$OH ratio of $62 \pm 3$, the latter suggesting clear fractionation effects when compared to the $^{12}$C/$^{13}$C-isotopologue ratios of other species.
\item We investigated the relative kinematics of the methanol lines in order to test for changes in $\mu$. We found clear evidence that the absorption velocities of the different transitions depend on the observation epoch, probably due to temporal variations in the system. We also confirmed a correlation between the absorption velocities of the different lines and their lower energy levels, which we interpreted as a signature of a temperature gradient in the absorbing gas. 
\item After taking these systematic effects into account by multivariate regression analysis, we did not find a significant correlation between the line bulk velocities and the \Kmu\ sensitivity factors to variations in $\mu$. We conclude that there is no evidence for changes in $\mu$ higher than $| \Dmu | = 3.6 \times 10^{-7}$ at 3$\sigma$ confidence level, at a look-back time of $\approx 7.5$~Gyr, which represents more than half the present age of the Universe.
\end{itemize}

This work illustrates the importance of addressing systematics effects to obtain a robust constraint on the cosmological invariance of fundamental constants. Future radio molecular absorption studies aimed at testing \Dmu\ below the $10^{-7}$ horizon, particularly those targeting \PKS1830, will have to be carefully designed.

\begin{acknowledgement}
  We thank the referee, Simon Ellingsen, for his constructive comments which improved the clarity of this work. This paper makes use of the following ALMA data: ADS/JAO.ALMA\#2018.1.00051.S. ALMA is a partnership of ESO (representing its member states), NSF (USA) and NINS (Japan), together with NRC (Canada) and NSC and ASIAA (Taiwan) and KASI (Republic of Korea), in cooperation with the Republic of Chile. The Joint ALMA Observatory is operated by ESO, AUI/NRAO and NAOJ. This research has made use of NASA's Astrophysics Data System.
  S.\,M. acknowledges support from Onsala Space Observatory for the provisioning of its facilities support. The Onsala Space Observatory national research infrastructure is funded through Swedish Research Council grant No 2017-00648. 
\end{acknowledgement}

\begin{appendix}

\section{Additional information} \label{app:line}

\begin{figure*}[h] \begin{center}
\includegraphics[width=12cm]{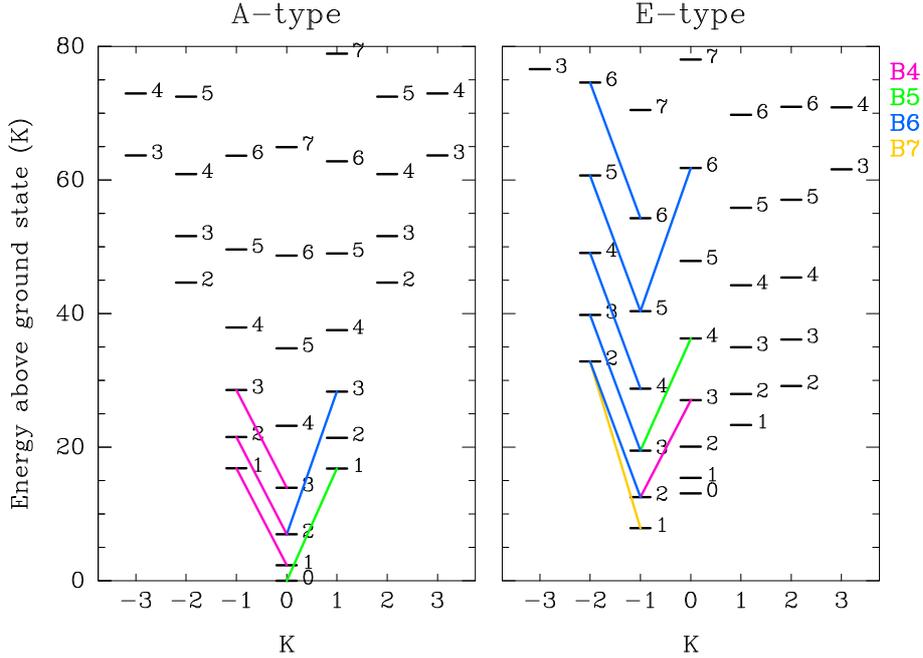}
\caption{Energy level diagram for the lower transitions of A-type (left) and E-type (right) methanol. The quantum numbers are $J$, as marked on the right of each level in the diagram, and $K$, as given on the x-axis. The transitions observed by us are indicated by the lines, for which the color-code corresponds to the different tunings. We note that the ground state level of E-CH$_3$OH lies 7.9~K above that of A-CH$_3$OH.}
\label{fig:Ediag}
\end{center} \end{figure*}

\begin{figure}[h] \begin{center}
\includegraphics[width=8.8cm]{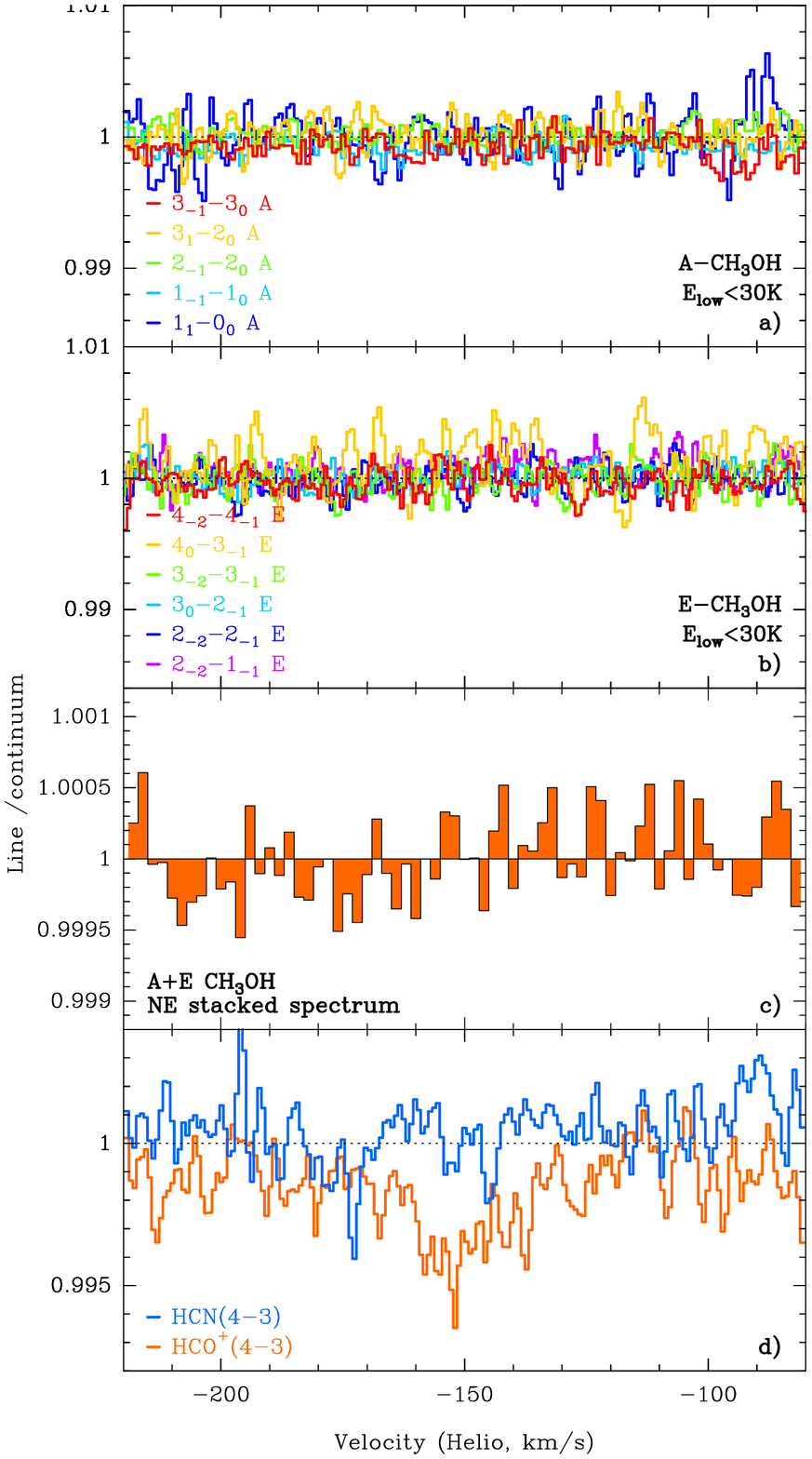}
\caption{(a,b) ALMA spectra of A/E-CH$_3$OH transitions observed toward the northeast image of \PKS1830. (c) Stacked spectrum, resulting from the weighted average of all A and E transitions with $\Elow<30$~K, regridded to a velocity resolution of 2~\kms. (d) For reference: HCO$^+$(4-3) and HCN(4-3) absorption spectra toward the northeast image.}
\label{fig:specNE}
\end{center} \end{figure}

The spectroscopic parameters of the submillimeter methanol lines observed in our campaign are given in Table~\ref{tab:line-methanol}. Their energy levels are plotted in Fig.\,\ref{fig:Ediag} for A-CH$_3$OH and E-CH$_3$OH separately. We note that the ground state of E-CH$_3$OH lies 7.90~K above that of A-CH$_3$OH.
The methanol spectra toward the northeast image of \PKS1830\ are shown in Fig.\,\ref{fig:specNE}, together with their stacked spectrum and with the spectra of HCO$^+$ and HCN $J=4-3$ transitions.

\begin{table*}[ht]
  \caption{Spectroscopic parameters of methanol lines observed by us.}
  \label{tab:line-methanol}
\begin{center} \begin{tabular}{clccrccc}
\hline \hline
Tuning & \multicolumn{1}{c}{Line} & Rest freq. $^{(a)}$ & Redshifted freq. & \multicolumn{1}{c}{\Elow $^{(b)}$} & $S_{ul}$ & $A_{ul}$ & \Kmu\ $^{(c)}$ \\
       & \multicolumn{1}{c}{($J_K$, type)}     & (MHz)     & ($z$=0.88582, GHz)           & \multicolumn{1}{c}{(K)} &  & ($10^{-4}$~s$^{-1}$) &     \\
\hline
B4 & CH$_3$OH $3_0$--$2_{-1}$ E & 302369.773 (0.012) & 160.339 & 12.54 & 0.49 & 0.47 & $-2.282$ \\
   & CH$_3$OH $1_{-1}$--$1_0$ A & 303366.921 (0.011) & 160.867 &  2.32 & 1.43 & 3.21 & $-1.904$ \\
   & CH$_3$OH $2_{-1}$--$2_0$ A & 304208.348 (0.011) & 161.314 &  6.96 & 2.38 & 3.23 & $-1.902$ \\
   & CH$_3$OH $3_{-1}$--$3_0$ A & 305473.491 (0.010) & 161.984 & 13.93 & 3.32 & 3.26 & $-1.898$ \\

   & $^{13}$CH$_3$OH $1_{-1}$--$1_{0}$ A & 303692.682 (0.050) & 161.040 &  2.27 & 1.43 & 3.22 & $-1.904$ \\
   & $^{13}$CH$_3$OH $2_{-1}$--$2_{0}$ A & 304494.300 (0.050) & 161.465 &  6.80 & 2.38 & 3.24 & $-1.902$ \\
   & $^{13}$CH$_3$OH $3_{-1}$--$3_{0}$ A & 305699.456 (0.050) & 162.104 & 13.59 & 3.32 & 3.27 & $-1.898$ \\
\hline
B5 & CH$_3$OH $4_0$--$3_{-1}$ E & 350687.662 (0.013) & 185.960 & 19.50 & 0.75 & 0.87 & $-2.105$ \\
   & CH$_3$OH $1_1$--$0_0$  A  & 350905.100 (0.011) & 186.076 &  0.00 & 0.95 & 3.32 & $-1.782$ \\
\hline
B6 & CH$_3$OH $2_{-2}$--$2_{-1}$ E & 423439.696 (0.012)  & 224.539 & 12.54 & 0.81 & 2.96 & $-1.292$ \\
   & CH$_3$OH $3_{-2}$--$3_{-1}$ E & 423468.647 (0.011)  & 224.554 & 19.50 & 1.42 & 3.73 & $-1.292$ \\
   & CH$_3$OH $4_{-2}$--$4_{-1}$ E & 423538.272 (0.011)  & 224.591 & 28.79 & 2.00 & 4.07 & $-1.291$ \\
   & CH$_3$OH $5_{-2}$--$5_{-1}$ E & 423675.185 (0.011)  & 224.664 & 40.39 & 2.56 & 4.27 & $-1.291$ \\
   & CH$_3$OH $6_{-2}$--$6_{-1}$ E & 423912.719 (0.010)  & 224.790 & 54.31 & 3.12 & 4.42 & $-1.290$ \\
   & CH$_3$OH $7_{-2}$--$7_{-1}$ E $^{(d)}$ & 424291.030 (0.010)  & 224.990 & 70.6 & 3.70 & 4.55 & $-1.289$ \\
   & CH$_3$OH $3_1$--$2_0$ A      & 445571.375 (0.011)    & 236.275 &  6.96 & 1.91 & 5.82 & $-1.615$ \\
   & CH$_3$OH $6_0$--$5_{-1}$ E    & 447118.364 (0.012) & 237.094        & 40.39 & 1.32 & 2.20 & $-1.865$ \\ 
\hline
B7 & CH$_3$OH $2_{-2}$--$1_{-1}$ E & 520179.054 (0.012) & 275.837 &  7.90 & 1.44 & 9.76 & $-1.237$ \\
\hline
\end{tabular}
\end{center}
\tablefoot{ $(a)$ Rest frequencies are taken from the Cologne Database for Molecular Spectroscopy \citep{CDMS01,CDMS05,CDMS16}, based on the works by \cite{xu08} for CH$_3$OH and \cite{xu97} for $^{13}$CH$_3$OH.
$(b)$ Energy of the lower level, with respect to the ground state of A-CH$_3$OH. For E-CH$_3$OH, the ground state lies 7.90~K above that of A-CH$_3$OH. This value needs to be subtracted to obtain the absolute \Elow\ values for E-CH$_3$OH.   
$(c)$ Sensitivity coefficients to a varying $\mu$, taken from \cite{jan11a} (see also Appendix~\ref{sec:Kmu}).
$(d)$ Not detected.}
\end{table*}

\section{Covering factor} \label{app:fc}

The covering factor, \fc, is the geometrical parameter that describes how much the absorbing material covers the background illuminating source ($I_{bg}$). The absorption spectrum $I_{abs}$ is connected to the opacity of the obscuring material via
\begin{equation}
I_{abs} = I_{bg} \times \left [ 1-\fc \times (1- \exp{(-\tau)}) \right ].
\end{equation}
It is easy to see that in the case of an optically thin line $(\tau << 1)$:
\begin{equation} \label{eq:degeneracy}
I_{abs} \sim I_{bg} \times (1-\fc \times \tau),
\end{equation}
and hence that there is an observational degeneracy between \fc\ and the true opacity.

How can we determine \fc? In the case of a heavily saturated line ($\tau >> 1$), $I_{abs}$ tends toward $I_{bg} \times (1-\fc)$, and \fc\ can then be measured directly on the flat-bottom shape of the spectrum around the line center. For species with hyperfine structure that are spectrally resolved with a good S/N, \fc\ and the true opacities can be determined under the assumption that the intensities of the different hyperfine components follow their statistical weights. When several lines with a spread in opacity are observed, then again, \fc\ can be fit with the assumption of a common line profile, provided a good enough S/N of the data.

We note that in general, and in particular in the case of \PKS1830, \fc\ could vary in time and frequency due to potential time evolution and chromatic morphology of the background source (see, e.g., \citealt{mar13}), as well as with line excitation, in the case of temperature gradients in the absorbing gas.

\section{Anomalous methanol absorption} \label{sec:AntiInversion}

In thermodynamic equilibrium, the relative population $n_i$ of the different energy levels $E_i$ is described by the Boltzmann equation:
\begin{equation}
\frac{n_u}{n_l} = \frac{g_u}{g_l} \exp{\left ( \frac{-(E_u-E_l)}{k_B \Tex} \right ),}
\end{equation}
where $u$ and $l$ denote an upper and lower level, respectively, $g_i$ is the level degeneracy, $k_B$, the Boltzmann constant, and \Tex, the excitation temperature. Under some circumstances in the ISM, however, the level populations can deviate from their equilibrium values. For example, when external radiative (e.g., strong infrared field) or collisional (e.g., high density and high kinetic temperature) pumping occurs, the upper energy levels of some transitions may become overpopulated ($n_u/g_u > n_l/g_l$), the so-called population inversion, potentially triggering intense maser emission (e.g., for OH, H$_2$O, and CH$_3$OH molecules). On the other hand, in regions with low temperature dominated by collisional excitation, the populations of some levels may become anti-inverted, with overpopulation of the lower energy level, depending on the balance between collisional rates and radiative decays. The excitation temperature may even become lower than the CMB temperature, and the molecule would then be seen in absorption against the CMB. This was observed for formaldehyde lines (e.g., toward Galactic dark clouds, \citealt{pal69}; see also \citealt{zei10}, and their discussion regarding the $z=0.68$ absorber toward B\,0218+357) and also for methanol lines (e.g., \citealt{wal88, pan08}).

\begin{figure}[h] \begin{center}
\includegraphics[width=8.8cm]{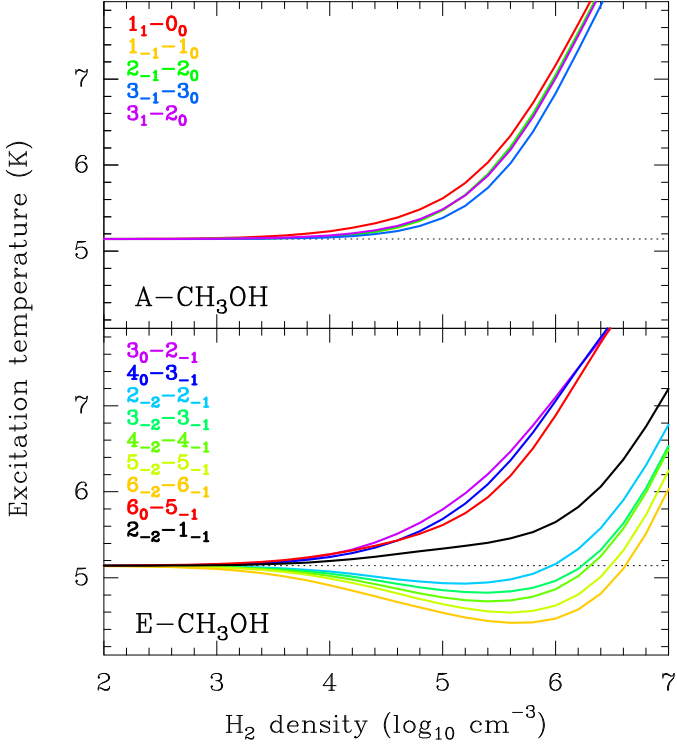}
\caption{Evolution of the excitation temperature with the H$_2$ density for methanol transitions in our survey, obtained from RADEX. We fixed a column density of $1 \times 10^{14}$~\cmsq, a FWHM of 1~\kms, and a kinetic temperature of 10~K. The CMB temperature at $z=0.89$ is marked by the dashed line.}
\label{fig:tex}
\end{center} \end{figure}

In Fig.\,\ref{fig:tex}, we show the evolution of the excitation temperature with volume density for all methanol transitions in our survey. At low density, $\nH2 \lesssim 10^3$~\ccm, \Tex\ stays coupled to the CMB temperature ($\Tcmb = 5.14$~K at $z=0.89$). At higher density, the collisions progressively raise \Tex until it eventually converges to \Tkin. However, for E-CH$_3$OH, all the $J_{-2}-J_{-1}$ transitions in our survey (J=2--6) show a decrease of \Tex\ below \Tcmb\ before they rise again beyond $\nH2 > 10^{6-7}$~\ccm. Under these conditions, the lines would be seen in absorption against the CMB, even if there were no background quasar emission.

\begin{figure}[h] \begin{center}
\includegraphics[width=8.8cm]{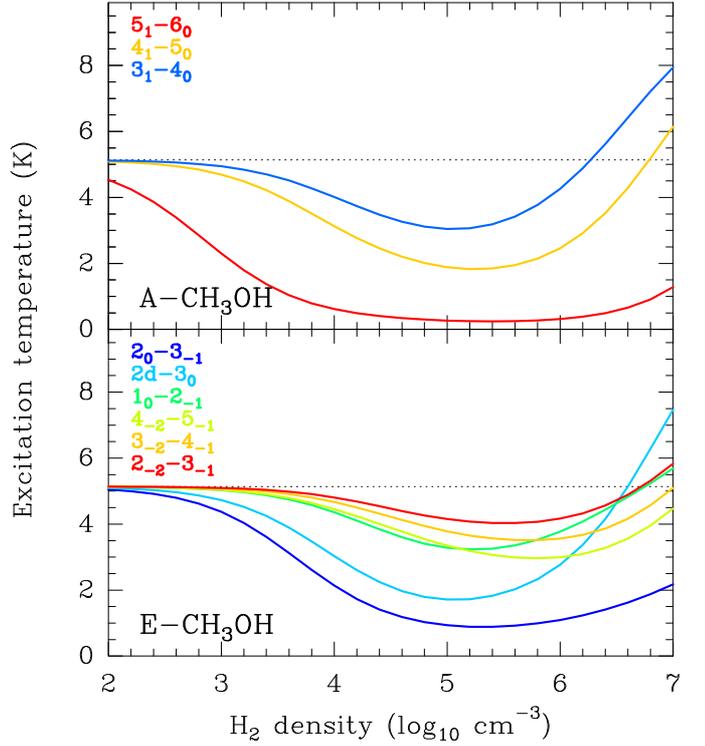}
\caption{Evolution of the excitation temperature with the H$_2$ density for methanol transitions harboring the maximum anti-inversion effect between 1 and 1000~GHz. The settings are the same as in Fig.\,\ref{fig:tex}. The CMB temperature at $z=0.89$ is marked by dashed lines.}
\label{fig:MaxAntiInversion}
\end{center} \end{figure}

Using RADEX, we further searched for methanol transitions between 1 and 1000~GHz possibly showing a large anti-inversion effect. We set a column density of $1 \times 10^{14}$~\cmsq\ for both A- and E-CH$_3$OH, a FWHM of 1~\kms, $\Tkin = 10$~K, and we fixed $\nH2 = 3 \times 10^5$~\ccm, since it is in the density regime where lines in our survey present the largest negative difference between \Tex\ and \Tcmb. From this first RADEX run, we selected the lines with $\Delta T = \Tcmb - \Tex > 1$~K and with peak opacity $\tau > 0.01$. Fig.\,\ref{fig:MaxAntiInversion} shows the evolution of \Tex\ with \nH2\ for all these selected lines. We find the maximum anti-inversion for the $5_1-6_0$ line of A-CH$_3$OH, at a rest frequency of 6.7~GHz, for which \Tex\ goes below 0.5~K for densities between $10^4$ and $10^6$~\ccm. It reaches $\Tex \sim 0.25$~K at its minimum. This line is associated with strong interstellar maser emission \citep{men91}, and it is thus prone to inversion or anti-inversion depending on the excitation conditions. As mentioned above, anti-inversion, leading to enhanced absorption, has indeed been observed in this line \citep{pan08}. For E-CH$_3$OH, the strongest anti-inversion is attained by the $2_0-3_{-1}$ line at rest frequency 12.2~GHz, first observed by \cite{wal88}. It is also known as a strong masing line \citep{bat87}. For this line,  $\Tex$ drops to a minimum value of $\sim 1$~K under the assumed physical conditions.

\section{\Kmu\ methanol sensitivity coefficients to $\mu$-variations} \label{sec:Kmu}

The values of the \Kmu\ sensitivity coefficients in Eq.\,(\ref{eq:mu}) need to be estimated by solving the torsional-rotational Hamiltonian of methanol as an internal rotor molecule. Using a simplified model with only six spectroscopic constants, approximate \Kmu\ were derived independently by \cite{jan11a} and \cite{lev11}.

 Here, we adopt the \Kmu\ values from \cite{jan11a} that were computed using a more sophisticated approach based on the BELGI representation of the full level structure of methanol, including the entire basis set of known transitions in this species, to high accuracy \citep{hou94}. The full model delivered accurate values for some 119 molecular constants representing the entire spectrum. It is known how these molecular constants depend on the reduced mass of the molecule and on the proton-electron mass ratio $\mu$, each with a certain power law. Values of \Kmu\ were then computed by taking the derivative of all transition frequencies with respect to $\mu$ at the actual value of the transition frequencies for $^{12}$CH$_3$OH. This analysis delivered \Kmu\ values that were surprisingly close to the values obtained from the very simplified models based on a set of six molecular constants pursued by \cite{jan11a} and by \cite{lev11}.
 
 In view of the fact that not all correlations among the 119 molecular constants in the BELGI model were investigated, \cite{jan11a} defined conservative estimates on the accuracy of the \Kmu, stating that they must be smaller than 5\%. The consistency between the full and the approximate models (the latter with considerably fewer parameters) indicates that the actual uncertainties should be less than 5\%.

The approach based on the BELGI analysis of the methanol molecule allowed \cite{jan11a} to also compute the \Kmu\ values for the $^{13}$CH$_3$OH isotopic species, with the derivatives of the transition frequencies with respect to $\mu$ taken at the frequencies of the $^{13}$C-isotopologue. For some specific transitions, and for transitions with high \Kmu\ values, large differences were found between the isotopologues, but for the transitions with small \Kmu\ values these appear to be similar and isotope independent. That holds for all transitions probed in the present study.

\end{appendix}

\end{document}